\documentclass[aps,pra,twocolumn,amsmath,amssymb,nofootinbib,superscriptaddress]{revtex4}

\usepackage[usenames, dvipsnames]{color}
\usepackage{graphicx}
\usepackage{epstopdf}
\usepackage{hyperref}
\usepackage{cleveref}
\usepackage{lpic}
\usepackage{blkarray}
\usepackage{array}
\usepackage{comment}
\usepackage{amsmath}

\newcommand{\ket}[1]{|#1\rangle}
\newcommand{\fig}[1]{Fig.~\ref{#1}}

\newcommand{\transpose}{^\mathsf{T}}

\newcommand{\tmsA}[1]{#1}
\newcommand{\tmsB}[1]{#1}
\newcommand{\mat}[1]{\underline{\underline{#1}}}
\newcommand{\ISF}{\text{ISF}}
\newcommand{\matconj}[1]{\underline{\underline{\widetilde{#1}}}}

\newcolumntype{L}[1]{>{\raggedright\let\newline\\\arraybackslash\hspace{0pt}}m{#1}}
\newcolumntype{T}[1]{>{\centering\let\newline\\\arraybackslash\hspace{0pt}}c{#1}}
\newcolumntype{C}[1]{>{\centering\let\newline\\\arraybackslash\hspace{0pt}}m{#1}}
\newcolumntype{R}[1]{>{\raggedleft\let\newline\\\arraybackslash\hspace{0pt}}m{#1}}

\usepackage{enumitem}
\newlist{steps}{enumerate}{1}
\setlist[steps, 1]{label = Step \arabic*:}

\graphicspath{{./figures/}}

\begin{document}

\title{Decoding Schemes for Foliated Sparse Quantum Error Correcting Codes}

\author{A. Bolt}
\affiliation{ARC Centre for Engineered Quantum System, Department of Physics, University of Queensland, Brisbane, QLD 4072, Australia}
\author{\tmsB{D. Poulin}}
\affiliation{D\'epartement de Physique, Universit\'e de Sherbrooke, Qu\'ebec, Canada}
\author{T. M. Stace} \email[]{stace@physics.uq.edu.au}
\affiliation{ARC Centre for Engineered Quantum System, Department of Physics, University of Queensland, Brisbane, QLD 4072, Australia}
\date{\today}
\frenchspacing

\begin{abstract}

Foliated quantum codes are a resource for fault-tolerant measurement-based quantum error correction for quantum repeaters and for quantum computation. They represent a general approach to  integrating a range of possible quantum error correcting codes into larger fault-tolerant networks. Here we present an efficient heuristic decoding scheme for foliated quantum codes, based on message passing between primal and dual code `sheets'.   We test this decoder on two different families of sparse quantum error correcting code: turbo codes and bicycle codes, and show reasonably high numerical performance thresholds. We also present a construction schedule for building such code states.
 
\end{abstract}

\maketitle

Quantum information processing (QIP) requires that the computational process must be performed with high fidelity.  In a noisy environment this will require quantum error correction (QEC)~\cite{shor_1995,steane_1996}. Depending on the computational model, this noise manifests in different ways.  A conceptually and technologically important step in the project to build quantum computers was the observation by Raussendorf et al.~\cite{raussendorf_2001,raussendorf_2003}  that highly-entangled \emph{cluster states}, are universal resource states with which to perform a quantum computation.  In cluster-state-based computation, the computation is driven forward by a series of measurements. 

Subsequently, Raussendorf et al.~\cite{raussendorf_2005,raussendorf_2006,raussendorf_2007a,raussendorf_2007b} proposed a method of \emph{fault-tolerant} quantum computation utilising {cluster states}. In this scheme a 3D cluster state lattice is constructed, which can be viewed as a \emph{foliation} of Kitaev's surface code~\cite{kitaev_1997,dennis_2002}. Alternating \emph{sheets} within this foliated structure correspond to primal or dual surface codes. Measurements on the \emph{bulk} qubits generate correlations between corresponding logical qubits on the boundary  of the lattice.  In these schemes, errors are partially revealed through parity checks operators, which can be determined from the outcomes of single qubit measurements.

Raussendorf's 3D measurement-based computation scheme has proved important for the practical development of quantum computers \cite{PhysRevA.86.032324,martinis_2014,Riste:2015aa,Gambetta:2017aa}, due to its high fault-tolerant computational error thresholds $\lesssim 1\%$\cite{dennis_2002}.  Furthermore, the robustness to erasure errors \cite{stace_2010,barrett_2010,duclos_2010}  makes these schemes attractive for  various architectures, including optical networks \cite{rudolph2017optimistic}. This high threshold is a result of the underlying surface code, which itself has a high computational error threshold $\sim 10\%$\cite{kitaev_1997,dennis_2002,stace_2010,barrett_2010}. 

Fault-tolerant, measurement-based quantum computation is achieved, in part, by `braiding' defects within the foliated cluster, to generate subgroup of the Clifford group. By virtue of their topologically protected nature, braiding operations can be made extremely robust, and so can be used to distill magic states.  Together, these resources allows for universal quantum computation~\cite{bravyi_2005,raussendorf_2007a,raussendorf_2007b}.

In an earlier paper, we showed that all Calderbank-Steane-Shor (CSS) codes can be \emph{clusterised} using a larger cluster state resource~\cite{bolt_2016}. These cluster state codes can be \emph{foliated} as a generalisation of Raussendorf's 3D lattice~\cite{bolt_2016}.  In that work we also demonstrated the performance of a turbo code with a heuristic decoder that we have developed.

 In this paper we present a detailed description of the decoding algorithm, and we apply the decoder to two classes of foliated CSS codes:  serial turbo codes \cite{berrou_1993,mceliece_1998,poulin_2009,utby_2006}, and bicycle codes. In contrast to the surface code, these code families have finite rate.  This allows for a much lower overhead of physical qubits to encoded qubits as the size of the code is increased, as compared to surface codes. In both cases, the  decoder on the complete foliated cluster state uses a soft-input soft-output (SISO) decoder within each sheet of the code  as a subroutine, followed by an exchange of marginal information between neighbouring primal and dual code sheets.  Iterating these steps yields an error pattern consistent with the error syndrome.

We present Monte Carlo simulations of the error-correcting performance using this decoder, assuming independently distributed Pauli $X$ and $Z$ errors.  We analyse the code performance in terms of both the Bit Error Rate (BER) and Word Error Rate (WER).  Our numerical results, simulating uncorrelated Pauli noise errors, indicate that the codes exhibit reasonably high \mbox{(pseudo-)thresholds} in the order of a few percent.  

In section~\ref{sec:clusterised_codes} we review the clusterization of CSS codes. Section~\ref{sec:foliated_codes} reviews the foliation of clusterised codes and presents a general decoding approach. In sections~\ref{sec:convolutional_trellis_construction}, we review the construction of decoding trellises for convolutional codes, which are a resource for the convolutional decoding. In section~\ref{sec:fol_conv_trellises} we develop decoding trellises for clusterised convolutional codes within a larger foliated code. In section~\ref{sec:trellis_siso_algorithm} we present a decoding algorithm for foliated convolutional codes. This is a subroutine for the foliated turbo decoder. In section~\ref{sec:numerical_results} we present the decoding scheme for foliated turbo codes and display our numerical results of simulated trials. Section~\ref{sec:foliated_bicycle_codes} presents a decoding algorithm for foliated bicycle codes, and presents numerical results. In section~\ref{sec:turbo_code_arch} we analyse the construction of clusterised convolutional and turbo codes from an architectural viewpoint. We investigate fault-tolerant constructions of foliated turbo codes.

\section{Clusterised Codes}
\label{sec:clusterised_codes}

\tmsB{We begin by reiterating some definitions, in order to set notation for what follows.}

\begin{figure}[t]
\begin{center}
\includegraphics[width=\columnwidth]{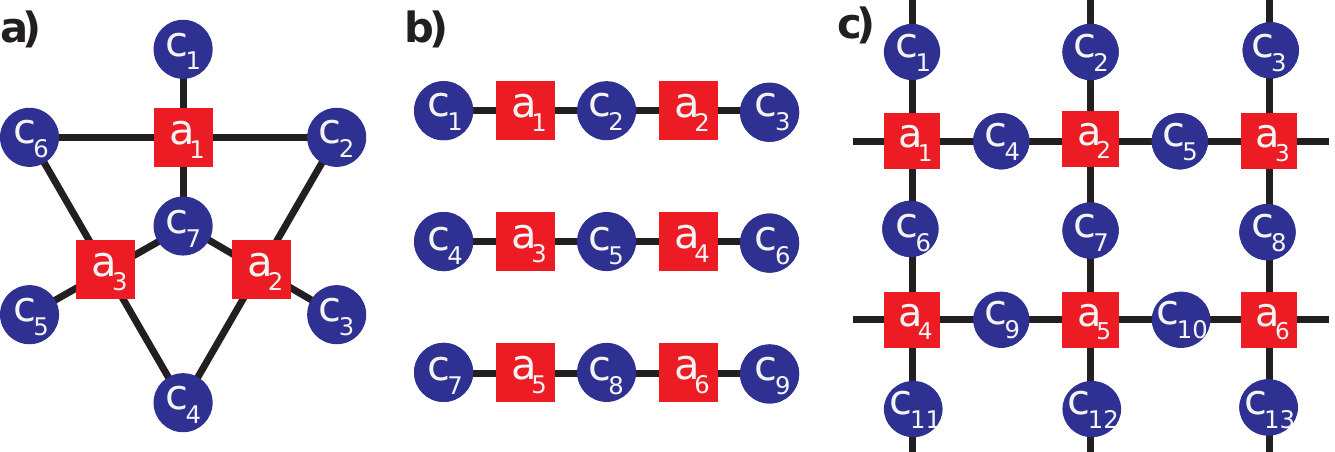}
\caption{Examples of progenitor clusters for clusterised CSS codes. {\bf a)} Clusterised Steane code. {\bf b)} Clusterised Shor code. 
 {\bf c)}  Clusterised surface code.  Code qubits (blue circles) are connected by cluster bonds (black lines) to ancilla qubits (red squares).  An $X$-basis measurement of  ancilla $a_k$  projects  neighbouring code qubits onto an eigenstate of \mbox{$\otimes_{\mathcal{N}_{a_k}} Z\in\mathcal{S}_Z$}.} \label{codeexamples}
\end{center}
\end{figure}

A \emph{cluster state} is defined on a collection of qubits located at the vertices of a graph \cite{raussendorf_2001,raussendorf_2003,briegel_2009}. A qubit at vertex $v$ carries with it an associated  cluster stabiliser $K_v=X_v(\otimes_{\mathcal{N}_v}Z)\equiv X_v Z_{\mathcal{N}_v}$, acting on it and  its neighbours,  ${\mathcal{N}_v}$.  The cluster state is the $+1$ eigenstate of the stabilisers \tmsB{$K_v$}.  \tmsB{For example, in \fig{codeexamples}(a), there is a cluster stabiliser $X_{a_1}Z_{c_1}Z_{c_2}Z_{c_6}Z_{c_7}$ associated to the ancilla qubit $a_1$.  Operationally, such a state can be produced with single and two qubit gates: each qubit is prepared in a $+1$ eigenstate of $X$, and then \textsc{c-phase} gates are applied to pairs of qubits that share an edge in the graph, e.g.\ in \fig{codeexamples}a, between qubit $c_2$ and its graph neighbours $a_1$ and $a_2$.}

\tmsB{A \emph{stabilser quantum code} is defined by the code stabilisers, $S$, which are a set of mutually commuting hermitian operators, whose simultaneous $+1$ eigenstates define valid code states.  A CSS code is one for which the  generators for $S$ can be partitioned into  a set of generators for $X$-like stabilisers, $\mathcal{S}_X$, which are products of Pauli $X$ operators acting on subsets of the code qubits,  and a set of generators for $Z$-like stabilisers, $\mathcal{S}_Z$, which are products of Pauli $Z$ operators, i.e.\ $S=\langle \mathcal{S}_X \cup \mathcal{S}_Z\rangle$.}

 An $[[n,k,d]]$ CSS code can be generated from  a larger \emph{progenitor} cluster  state \cite{bolt_2016}.   The progenitor cluster is the cluster state associated with the Tanner graph of $\mathcal{S}_Z$ {\cite{tanner_1981}}, i.e.\ a bipartite graph $G=(V,E)$ whose vertices $V$ are labeled as code qubits $c$, or ancilla qubits $a$. Each ancilla qubit $a$ is associated to a stabiliser  $Z_{\mathcal{N}_{a}}\in\mathcal{S}_Z$, so that $|V|=n+|\mathcal{S}_Z|$. $E$ contains the graph edge $(c,a)$  if $[H_Z]_{c,a} = 1$, where $H_Z$ is the parity check matrix.  The logical $X$ codestate of the CSS codes is obtained by measuring the ancilla qubits of the progenitor cluster state in the $X$ basis \cite{bolt_2016}.

Examples of clusterised CSS codes are shown in \fig{codeexamples}. These are the Steane 7 qubit code, 9 qubit Shor code and a 13 qubit surface code with $Z$ stabilisers generated by
\begin{align}
&\mathcal{S}_Z^{\text{Steane}} &\hspace{-3.5mm}=& \left\{ Z_{c_1}Z_{c_2}Z_{c_6}Z_{c_7},\, Z_{c_2}Z_{c_3}Z_{c_4}Z_{c_7},\, Z_{c_4}Z_{c_5}Z_{c_6}Z_{c_7}    \right\}\!,\nonumber\\
&\mathcal{S}_Z^{\text{Shor}} &\hspace{-3.5mm}=& \left\{ \begin{array}{l} Z_{c_1}Z_{c_2}, \,Z_{c_2}Z_{c_3}, \,Z_{c_4}Z_{c_5},\\Z_{c_5}Z_{c_6},\, Z_{c_7}Z_{c_8},\, Z_{c_8}Z_{c_9} \end{array}   \right\}\!,\label{stabilisers}\\
&\mathcal{S}_Z^{\text{Surf.}} &\hspace{-3.5mm}=& \left\{\begin{array}{l} Z_{c_1}Z_{c_4}Z_{c_6},\, Z_{c_2}Z_{c_4}Z_{c_5}Z_{c_7}, \, Z_{c_3}Z_{c_5}Z_{c_8},\\
 Z_{c_6}Z_{c_9}Z_{c_{11}}, \,Z_{c_7}Z_{c_9}Z_{c_{10}}Z_{c_{12}}, \,Z_{c_8}Z_{c_{10}}Z_{c_{13}}\end{array}\nonumber    \right\}\!.
\end{align}
For each of the $Z$-like stabilsers of a code, an ancillary qubit (red squares) is built into a cluster fragment with the associated code qubits (blue circles) in the stabiliser. For instance in the figure each ancillary qubit, $a_i$, is associated with the $i$th stabilizer element of $\mathcal{S}_Z$. This construction holds for all CSS codes~\cite{bolt_2016}.

CSS codes detect $X$ and $Z$ errors independently.  Each error type may be corrected independently, although this disregards potentially useful correlations, if such exist. We will assume independent $X$ and $Z$ Pauli noise, and in what follows, we describe the process for detecting and correcting $Z$ errors using $\mathcal{S}_X$ syndrome information.  The dual problem,  using $\mathcal{S}_Z$ syndrome information  to correct $X$ errors proceeds in exact analogy.

\tmsB{We note that for every CSS code, there is a \emph{dual} CSS code. The dual code is generated by exchanging $X$ and $Z$ operators in the stabilisers and logical operators.  That is an $X$-like stabliser in the primal code transforms to a $Z$-like stabiliser in the dual code, and vice versa.  Following the prescription above, a dual CSS code also has a progenitor cluster state, i.e.\ the dual code can also be clusterized in the same way.  If the code is self dual (e.g.\ the Steane code), then the corresponding primal and dual cluster states are identical.}

\section{Foliated Codes}
\label{sec:foliated_codes}

In this section we review the general construction of foliated codes as an extension of Raussendorf's 3D cluster state construction~\cite{raussendorf_2006,raussendorf_2007a,raussendorf_2005,raussendorf_2007b,bolt_2016}. Section~\ref{sec:foliated_construction} covers the generation of foliated codes from the cluster state resources in section~\ref{sec:clusterised_codes} and section~\ref{sec:foliated_decoding_approach} outlines a heuristic decoding approach which is suitable for general CSS codes. Specific implementations for convolutional, turbo and bicycle codes, which are all examples of  low-density parity check (LDPC) codes, are covered in later sections.%

\subsection{Foliated Code Construction}
\label{sec:foliated_construction}

\begin{figure}[t]
\begin{center}
\includegraphics[trim=0cm 0cm 0cm 0cm, clip=true,width=.9\columnwidth]{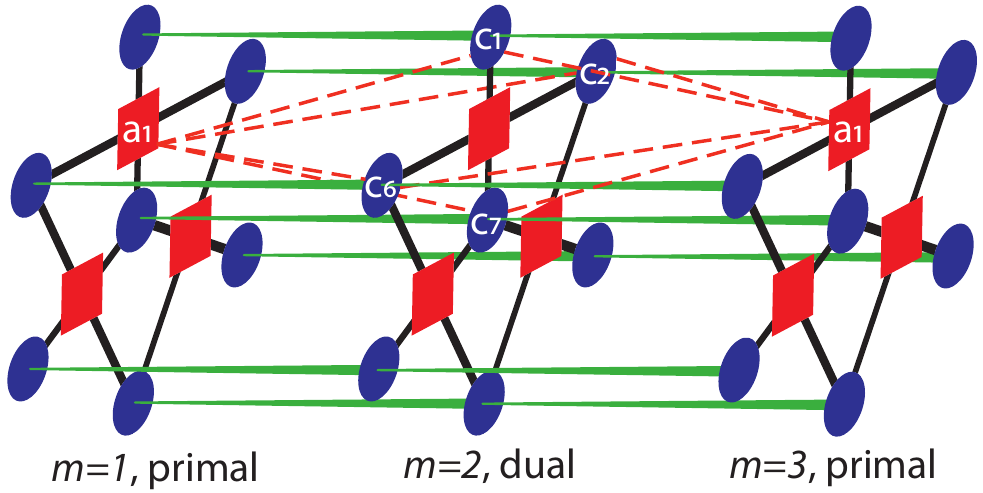}
\caption{A foliated Steane code with $L=3$ sheets. Code qubits (blue circles) share \tmsB{\textsc{c-phase}} cluster bonds (thick lines) with ancilla qubits (red squares) in the same sheet, and with code qubits in adjacent sheets (green streaked lines). The Steane code is self-dual, so that primal and dual cluster sheets are identical.  In this example, the end faces are indexed by $m=1$ and $m=L=3$.  Bulk qubits include all qubits in the $m=2$ sheet, and the ancialla qubits in the end faces. The product of cluster stabilisers centred on the labelled qubits \tmsB{(connected by dashed red lines) produces the parity check operator $\hat P_{1,m=2}$ in \cref{parityexample}}.
}
 \label{foliatedexample}
  \label{convex}
\end{center}
\end{figure}

The general foliated construction consists of alternating sheets of primal and dual clusterised codes as defined in \cref{sec:clusterised_codes}, which are `stacked' together \cite{bolt_2016}. 
\tmsB{The stacking of code sheets simply amounts to the introduction of additional cluster bonds (i.e.\ \textsc{c-phase} gates) between corresponding code qubits on neighbouring sheets.  This is depicted  in \cref{foliatedexample} for a specific code example (the Steane code of \cref{codeexamples}a).}

\tmsB{Suppose a CSS code has an $X$-like stabiliser, $\hat{s}$, with support on code qubits $c_{h_j}$, indicated  by the  support vector $\vec{h}=\{h_1,h_2,...\}$, i.e.\ $\hat{s}=X_{c_{h_1}}\otimes X_{c_{h_2}}...\equiv X_{c_{\vec{h}}}$.  In the foliated construction, there is a corresponding parity check operator centred on sheet $m$, given by}
\begin{align}
\hat P_{c_{\vec{h},m}}\equiv  X_{a_{\vec{h},m-1}}X_{c_{\vec{h},m}}X_{a_{\vec{h},m+1}},
\label{eq:foliatedparitycheck}
\end{align}
where $a_{\vec{h},m\pm 1}$ is the ancilla qubit associated with the dual code stabiliser $Z_{c_{\vec{h},m\pm 1}}$ acting on sheet $m\pm 1$.

\tmsB{ Given that each code sheet is derived from an underlying CSS code with logical operators $X_{\mathcal{L}}$ and $Z_{\mathcal{L}}$, we can define corresponding logical operators within each code sheet, $X_{\mathcal{L},m}$ and $Z_{\mathcal{L},m}$.    After the code is foliated, the logical operators in each code sheet commute with the  parity check operators \cite{bolt_2016}, i.e.\ $[\hat P_{c_{\vec{h},m}},Z_{\mathcal{L},m}]=0$.}

\tmsB{A reason for considering this construction is the observation that the foliated code cluster state provides a resource for error tolerant entanglement sharing.  This generalises one of the major insights of \citet{raussendorf_2005}, in which is was shown that after foliating $L$ surface code sheets, the resulting 3-dimensional cluster state (defined on a cubic lattice) served as a resource for fault-tolerantly creating an entangled Bell pair of  surface code logical qubits between the first and last sheet (labelled by the index $m=1$ and $L$).  Starting from the 3-dimensional foliated surface code cluster, this long range entanglement is generated by  measuring each of the \emph{bulk physical qubits} (i.e.\  all ancilla qubits, and all code qubits in sheets $m=2$ to $L-1$) in the $X$ basis.  
Formally, this is shown by noting that after the bulk qubit measurements, the remaining physical qubits (which are confined to sheets 1 and $L$) are stabilised by the operators $X_{\mathcal{L},1}\otimes X_{\mathcal{L},L}$ and $Z_{\mathcal{L},1}\otimes Z_{\mathcal{L},L}$ ~\cite{raussendorf_2005}, up to Pauli frame corrections that depend on the specific measurement outcomes on the bulk qubits.  The underlying surface code makes the protocol described therein robust against pauli errors on the bulk qubits.}  

\tmsB{As discussed in \cite{bolt_2016}, this property  is respected for \emph{any} foliated CSS code.  That is, measurements on the bulk qubits project the  logical qubits encoded within the end sheets into an entangled logical state. This is verified by checking that $X_{\mathcal{L},1}\otimes X_{\mathcal{L},L}$ and $Z_{\mathcal{L},1}\otimes Z_{\mathcal{L},L}$ are in the stabiliser group of the cluster state after bulk measurements are completed.}

For an underlying $[[n,k,d]]$ code, there are weight-$d$ undetectable error chains on the foliated cluster, as in \cite{raussendorf_2006,raussendorf_2007a,raussendorf_2005,raussendorf_2007b,dennis_2002}.  Since the structure of the code in the direction of foliation is a simple repetition, it follows that the foliated cluster inherits the distance of \mbox{the underlying code.}

\tmsB{\fig{foliatedexample} shows an example of the cluster state for a foliated Steane code.  
Alternating sheets of the primal Steane code cluster state (shown in \fig{codeexamples}a) and its self-dual  are stacked together, with additional cluster bonds (green streaked lines) extending between corresponding code qubits in each sheet; operationally these correspond to \textsc{c-phase} gates between code qubits.  The Steane code is self-dual, so primal and dual sheets are identical. }

\tmsB{An example of a parity check operator centred on sheet $m=2$ is 
\begin{equation}
\hat P_{1,m=2}=X_{a_1,1}X_{c_1,2}X_{c_2,2}X_{c_6,2}X_{c_7,2}.X_{a_1,3},\label{parityexample}
\end{equation}
 is depicted in \fig{foliatedexample}.
For this example, there are two other parity check operators centred on sheet $m=2$, associated to the other ancilla qubits therein.  The logical operators for the Steane code pictured in \cref{codeexamples}a are $Z_\mathcal{L}^{\text{Steane}}=Z_{c_1}Z_{c_2}Z_{c_3}$ and $X_\mathcal{L}^{\text{Steane}}=X_{c_1}X_{c_2}X_{c_3}$. By inspection, the parity check operator $\hat P_{1,m}$ commutes with $Z_{\mathcal{L},m}^{\text{Steane}}$ on sheet $m$.}

\tmsB{\fig{foliatedexample} also illustrates a minimal example of entanglement sharing between the end sheets of the foliated construction, for $L=3$ code sheets.  After the foliated cluster state is formed, $X$ measurements on the bulk qubits (all qubits in the dual sheet shown, and the ancilla qubits in the end primal sheets) leaves the remaining physical qubits (the code qubits in the primal end sheets)  stabilised by the operators $X^{\text{Steane}}_{\mathcal{L},1}\otimes X^{\text{Steane}}_{\mathcal{L},3}$ and $Z^{\text{Steane}}_{\mathcal{L},1}\otimes Z^{\text{Steane}}_{\mathcal{L},3}$.}

\tmsB{ Additional examples of the Shor code and surface code are presented in~\cite{bolt_2016}. }

 \subsection{Decoding Approach}
 
 \label{sec:foliated_decoding_approach}
 
 \begin{figure*}[t]
 \begin{center}
\includegraphics[width=0.95\textwidth]{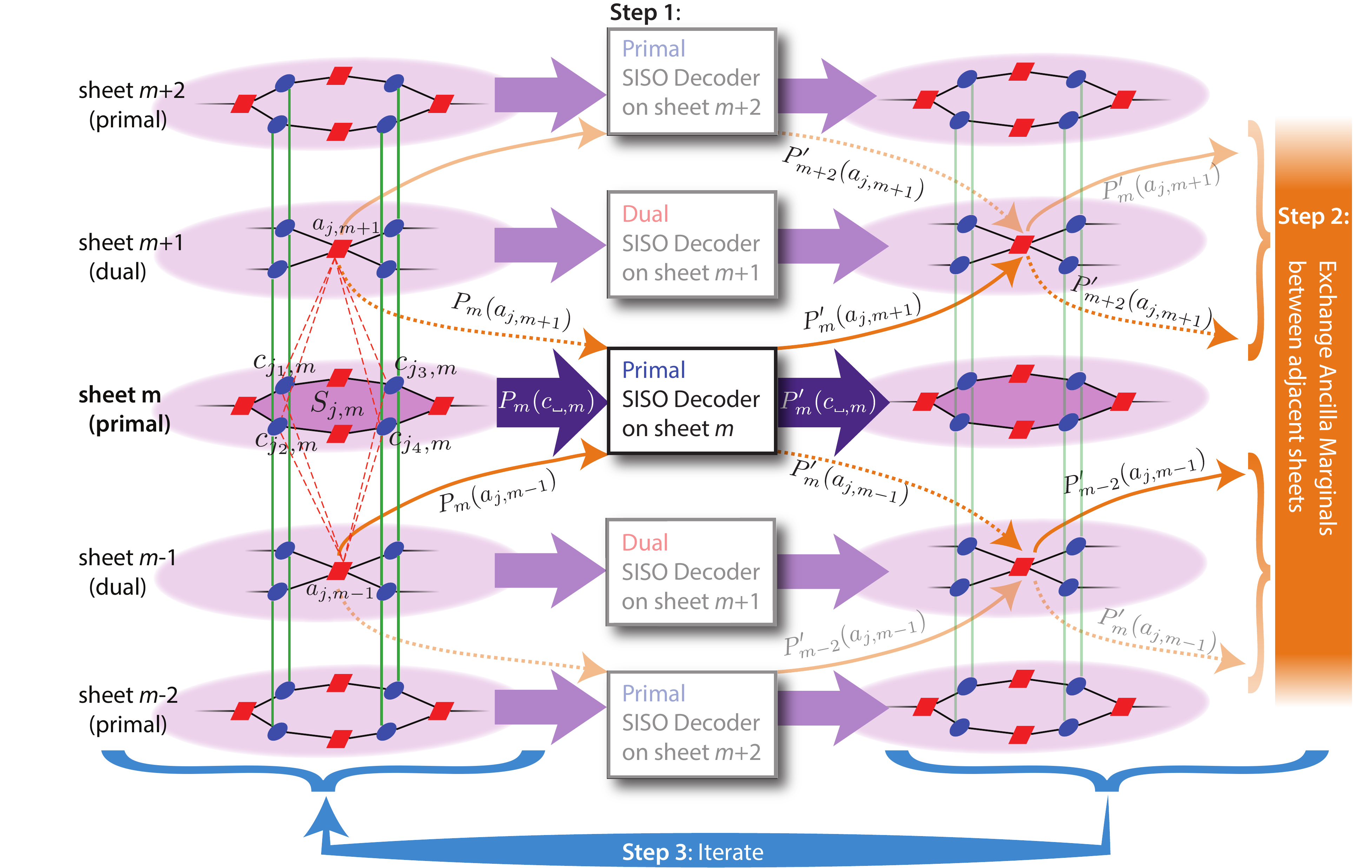}
\caption{\tmsB{A schematic flow of \emph{beliefs} (i.e.\ marginal probabilities) in the heuristic decoder for a generic foliated cluster code.  Code qubits are shown as circles, ancilla qubits as rectangles within each code sheet.  Cluster bonds are shown as black (green) lines within (between) code sheets (as in \cref{foliatedexample}).  The decoding sheet consists of code qubits $c_{{\textvisiblespace},m}$ in sheet $m$, and the ancilla in the adjacent sheets, $a_{{\textvisiblespace},m\pm1}$.  At the left, one stabiliser, $S_{j,m}$, with support on code qubits $c_{j_{1,...4},m}$ in sheet $m$  is shown (shaded), and forms a parity check operator with  the associated  ancilla qubits $a_{j,m\pm1}$ (connected by dashed red lines).  \textsc{\mbox{Step 1}}: on each decoding sheet $m$, the soft-input soft-output (SISO) decoder receives syndrome data, $\vec S_m$ and  input error marginals $P_m({c_{\textvisiblespace},m})$ on the code qubits (thick purple arrows)  and adjacent ancilla qubits $P_{m\pm1}({a_{\textvisiblespace},m})$ (thin orange arrows) to compute updated error marginals $P'_m$ for code and ancilla qubits.  This step can be processed in parallel across all sheets, since there are no cross-dependencies.  Note that after \textsc{Step  1} there are \emph{two} marginals associated to each adjacent ancilla: e.g.\ $P_m'(a_{j,m+1})$ (solid orange arrow emerging from the decoder) and $P_{m+2}'(a_{j,m+1})$ (dotted orange arrow emerging from the decoder), which are computed from decoders on sheets $m$ and $m+2$ respectively.  If these disagree, $P'_m(a_{j,m+1})\neq P'_{m+2}(a_{j,m+1})$, the value are exchanged in  \textsc{Step  2}.  After marginal exchange, the process is iterated (\textsc{Step 3}), until marginals converge.  After marginals on the ancilla have converged sufficiently, maximum likelihood decoding is performed within each code sheet (\textsc{Step 4}, not shown)}.
}
\label{fig:loop}
\end{center}
\end{figure*}
 

 Errors in the foliated cluster are detected by parity check operators: a $Z$ error will flip one or more parity checks, giving a non-trivial error syndrome for the foliated cluster.   Importantly, the parity check measurement outcomes can be inferred from products of single-qubit $X$ measurement outcomes. The syndrome information in then input into a decoder.  
 
While generic CSS codes may not be efficiently decoded, many exact or heuristic decoders are known for specific code constructions\cite{dennis_2002,ollivier_2003,poulin_2009,duclos_2010}. For the purpose of this paper, we assume that whichever code is chosen,  a practically useful decoder is known. \tmsB{In what follows, this decoder forms a subroutine that is called repeatedly in the decoding of the larger, foliated code.}
 
\tmsB{We note here that this is qualitatively different from the foliated surface code \cite{dennis_2002,raussendorf_2003}, for which the decoder does \emph{not} call the surface code decoder as a subroutine.  Instead}, the surface code decoders are \emph{generalised} to the foliated version. For instance, the minimum weight perfect matching decoder for the surface code can be  modified to the foliated case~\cite{dennis_2002,raussendorf_2005}, by replacing stabiliser defects in the 2D Kitaev lattice \cite{kitaev_2003}, with parity check defects in the 3D Raussendorf lattice \cite{dennis_2002,raussendorf_2003}, but still using perfect matching on the syndrome. However this is a peculiarity of the Raussendorf construction as a graph product code \cite{tillich_2009}, constructed from repeated graph products of a repetition code.

Here, we propose a general purpose, heuristic method of decoding foliated codes, which is based upon the  decoding of the underlying CSS codes in the presence of faulty syndrome extraction. If the decoder  for the underlying code is efficient then it will be efficient for the full foliated process. The foliated decoding algorithm is a heuristic method based on belief propagation (BP) that may work well in many cases~\cite{mceliece_1998,poulin_2008}. The overall foliated decoder calls a Soft-Input Soft-Output\footnote{\tmsB{`Soft' values indicate that the decoder accepts and returns probabilities for errors, as opposed to `hard' values which are binary allocations (`error' OR `no error') at each qubit. For example, perfect matching on the surface code is a hard decoder.}} (SISO) decoder for each of the underlying primal and dual  CSS codes, as a subroutine.  The SISO decoder calculates the probability of a Pauli error, $\sigma$, on qubit $q \in \{a_k,c_j \}$, $P(\sigma_q | \vec S_{\text{CSS}})$, given a physical error model and syndrome data for the CSS code, $\vec S_{\text{CSS}}$, which may itself be unreliable, due to errors on the ancilla qubits. Using such a decoder it is possible to assign a probability of failure to a parity check to account for errors on ancilla qubits, \mbox{$P(\sigma_a)=\sum_{\vec S_{\text{CSS}}}P(\sigma_a | \vec S_{\text{CSS}})P( \vec S_{\text{CSS}})$}.

For the foliated case, consider a parity check operator given by \cref{eq:foliatedparitycheck}. A non-trivial syndrome can arise because of errors on  code qubits $\vec{h}$  within  code sheet $m$, or due to errors on the corresponding ancilla qubits, $a_{\vec{h}}$ in adjacent dual sheets $m\pm 1$.
  
In the case where dual-sheet ancilla qubits are error-free the decoding problem reduces to a series of independent CSS decoders using perfect syndrome extraction. However, errors on the ancillas mean that the in-sheet syndrome is unreliable. To account for the dual-sheet ancilla errors we embed the CSS decoder in a \emph{belief propagation} (BP) routine, as iterated in the following scheme. \textsc{Steps} 1, 2 and 3 and illustrated in \cref{fig:loop}.

\begin{enumerate}
\setcounter{enumi}{-1}

\item A \emph{decoding sheet} centred on sheet $m$ is defined by a set of qubits, $q$, which contain the code qubits $c_{j,m}$ within the sheet and the ancilla qubits $a_{k,m\pm 1}$ on neighbouring sheets. The input to the decoder is the syndrome data $\vec S_m$ on sheet $m$, and a prior probability distribution, $P^{\textrm{pr}}_m(a_{k,m\pm 1})$, for the marginals on the ancilla qubits associated to the decoding sheet $m$.

\item The SISO decoder on sheet $m$ computes marginal error probabilities for the code qubits in the sheet, $P'_m(c_{j,m})$, assuming marginals on the ancilla in sheet $m\pm1$,  and updated marginals on the ancilla qubits $P'_m(a_{k,m})$.  This step can be parallelised across all sheets.

\item The assumed error model for ancilla qubits is updated using the results from step 1. \mbox{$P_m(a_{k,m\pm 1})\rightarrow P'_m(a_{k,m\pm 1}) = P_{m\pm 2}(a_{k,m\pm 1}|S_{m\pm 2})$}, where $P_{m\pm 2}(a_{k,m\pm 1})$ is the probability distribution found from a neighbouring decoding process. We refer to this update rule as  an \emph{exchange of marginals}. 

\item  Using the updated ancilla marginals from Step 2, we iterate back to Step 1 until  ancilla marginals converge sufficiently, i.e.\ $P_m(a_{k,m\pm 1})\approx P'_m(a_{k,m\pm 1}) $, in which case we proceed to Step 4. 

\item Use the \tmsB{converged} marginals computed on the code and ancilla qubits, $P_m(c_{j,m})$ and $P_m(a_{k,m\pm1})$, to calculate an error correction chain (parametrised by  error correction binary support vector, $\vec e$) with (approximately) maximum likelihood $P_m(\vec e |\vec S_m, P_m(c_{j,m}), P_m(a_{k,m\pm 1}))$, within each code sheet.  \tmsB{This could employ a hard decoder.}
\end{enumerate}

In sections~\ref{sec:convolutional_trellis_construction}-~\ref{sec:foliated_bicycle_codes} we describe specific soft decoding implementations for foliated convolutional, turbo, and LDPC bicycle codes. For convolutional and turbo codes we first introduce a trellis decoding framework, which is then adapted for use as a single layer decoder. This is then combined with the BP process above to generate a full decoding scheme. For bicycle codes we use belief propagation directly on the Tanner graph representation of the foliated code.

\section{Convolutional Trellis Construction}
\label{sec:convolutional_trellis_construction}
In this section we review the construction of trellises as a tool for SISO  decoding of convolutional codes~\cite{forney_1965,mceliece_1998,sadjadpour_2000,poulin_2009}
Section~\ref{sec:fol_conv_trellises} modifies this construction for use with single sheets of the foliated code, which is itself a subroutine in the full decoding of foliated convolutional codes.

Generators and stabilisers of convolutional codes are translations of some `seed' generators or stabilisers which act over a sequence of \emph{frames}.  \tmsB{Each frame labels  a contiguous block of $n$ qubits.} 
We assume here that the code has $\tau$ frames, labelled by a frame index, $t=1,...,\tau$.. For a classical rate $\frac{k}{n}$ code the generator matrix (for logical $Z$ operators) has the form:
\begin{align}
 \textit{frame: }  & \hspace{2.75mm}\begin{array}{C{3mm}C{5mm} C{5mm}C{5mm}C{5mm} C{5mm}C{5mm}C{5mm}}
$...$ & $t\!-\!1$ & $t$ & $t\!+\!1$ & $t\!+\!2 $& $...$ & &
 \end{array} \nonumber\\
\mathbf{G}\transpose    =& \left[ \begin{array}{C{3mm}C{5mm} C{5mm}C{5mm}C{5mm} C{5mm}C{5mm}C{5mm}}
$...$& $\!G^{(\!1\!)}$     & $\!\!G^{(\!2\!)}$     &$...$ & $\!G^{(\!\nu_g\!)}$ &$...$  &      &\\
& $...$     &$\! \!G^{(\!1\!)}$     & $\!G^{(\!2\!)}$ &$...   $&$ \!G^{(\!\nu_g\!)}\!$&$...$&\\
  &         & $...$    & $\!G^{(\!1\!)}\!$   & $\!G^{(\!2\!)}\!$     &$ ... $&$\!G^{(\!\nu_g\!)} $&$...$
\end{array}\right]_{{k}\tau\times n\tau}
\label{eq:fullgenerator}
\end{align}
where $G^{(i)}$ are binary-valued $k \times n$ sub-matrices.  We use the notation $A\transpose$ to denote the  transpose of the matrix $A$; bold face matrices indicate generators acting on the entire set of physical qubits. All other elements in $G$ are zero. Each $G^{(i)}$  acts on a single frame of $n$ physical bits, encoding $k$ logical bits. The code is built from translations of the sub-matrices $[G^{(1)},\hdots, G^{(\nu_g)} ]$. Each component, $G^{(j)}$, acts on a single {frame} of the code. The value of $\nu_g$ is the \emph{codeword memory} length, \tmsB{which counts the number of frames over which parity check operators have support.}

\tmsB{Later we will discuss a specific example of a convolutional code that illustrates this construction.  For concreteness, we preempt that example by reference to \cref{fig:marginal_exchange}a, each row of which depicts a convolutional code  with frames consisting of $n=3$ qubits (blue circles), and with parity check operators (red squares) extending over $\nu_g=3$ frames.  Note that in this example, each parity check operator has support on 6 of the qubits (indicated by thick black lines) within the $\nu_g=3$ contiguous frames.}

Similarly, we define the parity check generator matrix 
\begin{align}
\mathbf{H} \!\!=\! \left[ \begin{array} {c cc ccc c}
\hdots& H^{(1)}        &\hdots & H^{(\nu_h)} &\hdots  &      &\\
& \hdots     & H^{(1)}    &\hdots   & H^{(\nu_h)}&\hdots&\\
  &         & \hdots    & H^{(1)}       & \hdots &H^{(\nu_h)} &\hdots
\end{array}\right]_{{n_z}\tau\times n\tau},\label{eq:uncompactH}
\end{align}
where ${n_z}=|\mathcal{S}_Z|/\tau$ is the number of $Z$-like stablisers per frame. The value of $\nu_h$ is the \emph{parity check  memory} length. Typically $\nu_h$ and $\nu_g$ are of similar size.

Codeword generators are expressed in the form $\hat{g} = Z^{\otimes \vec{g}}$, where $\vec{g}\in \mathbb{Z}_2^{n\tau}$ is in the row space of $\mathbf{G}_Z$. We have introduced the notation \mbox{$Z^{\otimes \vec{v}} = Z_1^{v_1} \otimes Z_2^{v_2} \hdots$} with $v_j\in \mathbb{Z}_2$. Similarly the stabiliser generators formed from $\mathbf{H}_Z$ are $\hat{h} = Z^{\otimes \vec{h}}$, where $\vec{h}\in \mathbb{Z}_2^{n\tau}$ is in the row space of $\mathbf{H}_Z$.

The commutation relationships between generator and stabiliser matrices manifests as orthogonality conditions, i.e.\
\begin{align}
\left[\begin{array}{c}\mathbf{G}_Z\transpose\\\mathbf{H}_{Z}\\ \textbf{ISF}_Z \end{array} \right]\!\!\left[\begin{array}{lll}\mathbf{G}_X& \textbf{ISF}\transpose_X & \mathbf{H}_X\transpose \end{array} \right] &=\mathbb{I}_{n\tau\times n\tau },
\label{eq:commutation_conditions}
\end{align}
where we have introduced the \emph{Inverse Syndrome Formers} (ISFs). The ISF is useful  determining an initial, valid decoding pattern, known as a \emph{pure error}. \tmsB{Each row of $\textbf{ISF}_Z$ corresponds to an operator that   commutes with all $X$-like stabiliser generators, $\hat g\in \text{RowSpace}({\mathbf{G}_X})$, and with all but one of the parity check generators $\in \text{RowSpace}({\mathbf{H}_X\transpose})$; there are as many ISF generators as there are parity check generators. 
 The rows  of $\mathbf{G}_Z\transpose$ and $\mathbf{H}_{Z}$ form an orthonormal set, but do not fully span $\mathbb{Z}_2^{n\tau}$. The $\textbf{ISF}_Z$ sub-matrix can be computed  by finding an  orthonormal completion of the  rowspace of $\mathbf{G}_Z\transpose$ and $\mathbf{H}_{Z}$, e.g.\  using  Gram-Schmidt. The matrix $\textbf{ISF}_Z$ is not unique: any orthogonal completion of the basis that satisfies $\textbf{ISF}_{Z}\cdot{\mathbf{H}_X\transpose}=\mathbb{I}$ will suffice\footnote{i.e.\ $\textbf{ISF}_{Z}\transpose$ is a pseudo-inverse of $\mathbf{H}_{X}$, and vice-versa.}.  Similarly $\textbf{ISF}\transpose_X$ is generated from an orthonormal completion of the column space of $\mathbf{G}_X$ and $\mathbf{H}_X\transpose$.}

\tmsB{Suppose some (unknown)  pattern  $\vec\varepsilon\in \mathbb{Z}_2^{n\tau}$ of $Z$ errors gives rise to  a syndrome that is revealed by the $X$-like parity checks, i.e.\ $\vec S=\mathbf{H}_{X}\vec\varepsilon\in\mathbb{Z}_2^{n_x\tau}$.  We use $\textbf{ISF}_Z$ to generate a  pure error, $\hat E^0 = Z^{\otimes \vec{e}^{\,0}} \equiv Z^{\otimes(\textbf{ISF}_Z\transpose \cdot \vec S)}$, based only on the syndrome data.  The binary-valued support vector $\vec{e}^{\,0}\in \mathbb{Z}_2^{n\tau}$ corresponds to a possible error correction pattern that satisfies the syndrome $\vec S$, i.e.\ $\vec S=\mathbf{H}_{X} \vec\varepsilon=\mathbf{H}_{X}\vec{e}^{\,0}$.  Therefore  $\hat E^0$ is a \emph{valid} decoding pattern, but it is unlikely to be the most probable decoding, and is so is unlikely to robustly correct the original error. However it  defines a reference decoding from which the set of \emph{all} valid decodings, $\mathcal{E}$}, can be enumerated through 
\begin{align}
 \mathcal{E} &= \! \{ {\hat E}^0  \hat h   \hat g| \hat h\in \mathcal{S}_Z, \hat g\text{ is a code word}\} ,\label{eq:trellis_paths} \\
 &=  \!\{ \!Z^{\otimes  (\vec{e}^{\,0} + \vec{h} + \vec{g} )}|\vec h\!\in \!\text{RowSpace}(\mathbf{H}_Z\!), \vec g\in \text{RowSpace}(\mathbf{G}\transpose_Z\!)\},\nonumber
\end{align}
where we take linear combinations of rows of $\mathbf{H}_Z$ and $\mathbf{G}\transpose_Z$ over $\mathbb{Z}_2$. In what follows, we suppress the subscripts $X$ and $Z$.

A valid decoding of the syndrome data may be written as ${\hat E}^0Z^{\otimes \vec p}$, where $\vec p\equiv\vec h+\vec g\in \text{RowSpace}(\mathbf{H})\cup\text{RowSpace}(\mathbf{G}\transpose)\subset \mathbb{Z}_2^{\tau n}$. 
That is we define
\begin{align}
\vec p &=
\sum_{i=1}^k\sum_{j=1}^\tau l_{i,j}^{(g)}{\mathbf{G}}\transpose_{i+j}+\sum_{i=1}^{n_z}\sum_{j=1}^\tau l_{i,j}^{(h)}{\mathbf{H}}_{i+j},\label{pvec}
\end{align}
where $\mathbf{A}_i$ refers to the $i^\text{th}$ row of $\mathbf{A}$.  

 \tmsB{Relative to the (easily found) pure error support vector, $\vec{e}^{\,0}$,  the vector $\vec p$ parametrises all possible valid decoding through the coefficients $ l_{i,j}^{(g)}$ and $ l_{i,j}^{(g)}$.} A  good decoder will return optimal values of $l_{i,j}^{(g)}$ and $l_{i,j}^{(h)}$, corresponding to an element, $\hat{E}=Z^{\vec{p}+\vec{e}^{\,0}} \in \mathcal{E}$, that has a high likelihood of correcting the original error.  For low error rates, this amounts finding  a $\vec p$ that minimises the Hamming weight of $\vec p+\vec{e}^{\,0}$. 
 
\tmsB{For readers unfamiliar with this general construction, we provide a short example based on the 7-qubit Steane code in Appendix \ref{ISFexample}.}

\subsection{Seed Generators of Convolutional Codes}

Finding an optimal $\vec p$ by enumerating over all $2^{(k+n_z)\tau}$ possible binary values of $l_{i,j}^{(g)}$ and $l_{i,j}^{(h)}$ becomes computationally intractable  as the size of the code is increased. However the repeated structure of convolutional codes, in blocks of length $\nu n$, allows for a simplification, using a decoding trellis.  The trellis decoder reduces the search space to $\sim 2^{k\nu_g+n_z\nu_h} \tau$, so that the decoding is linear in the code size $\tau$, albeit with potentially large prefactor depending on the total memory lengths $\nu_g$ and $\nu_h$.

As the number of encoded bits increases, the size of the generator matrix increases. We therefore use a more compact representation using transfer functions, which are polynomials in the \emph{delay} operator, denoted by $D$. We  interpret $D$ as a discrete generator of frame shifts, so that $D^q$ represents a `delay' of $q$ frames.  We also define the inverse shift generator, $\tilde D$, which acts like a reverse translation such that $D^a\widetilde{D}^b = \delta_{ab}$. Detail of the construction are given in Appendix~\ref{ap:trans_func_notation}.

 Using this delay notation, we define the \emph{seed matrix} of a rate $\frac{k}{n}$ convolutional code in terms of the sub-matrices, $G^{(i)}$, in \cref{eq:fullgenerator}
\begin{align}
G\transpose(D) &\equiv G^{(1)}+D G^{(2)}+...+D^{\nu_g-1} G^{(\nu_g)},\nonumber\\
&\equiv \left[\begin{array}{cccc}
g_{11}(D) & g_{12}(D) & \hdots & g_{1n}(D)\\
\vdots &        &        & \vdots\\
g_{k1}(D) & g_{k2}(D) & \hdots & g_{kn}(D)
\end{array} \right]_{k\times n},
\end{align}
where $g_{ij}(D)$ is a polynomial in $D$, defined by
\begin{align}
g_{ij}(D) &= \sum\limits_{q = 1}^{\nu_g} D^{q-1}G^{(q)}_{ij}.
\end{align}
The utility of the delay operator notation is that (1) it enables us to write $G\transpose(D)$ in terms of a matrix that is independent of the number of frames, $\tau$, and (2) it sets the degree of the polynomial entries.
%
\begin{figure}[t]
\includegraphics{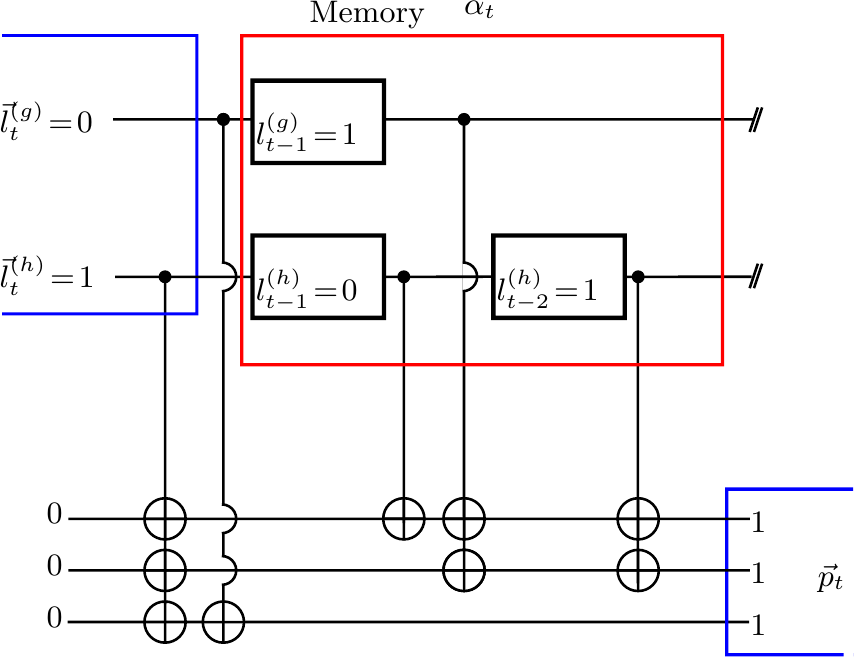}
\caption{A schematic circuit for computing $\vec p_t$ in \cref{UP} for a single frame, $t$, based on the the $\tfrac{k}{n}=\tfrac{1}{3}$ convolutional code defined in \cref{eq:example_conv}. The circuit receives values of $\vec{l}_t$ for  frame $t$, and returns a $\vec p_t$, which depends the recent history of $\vec{l}$, which is stored in memory elements  (black boxes). For an input of $\vec{l}_t = \{0;1\}$ and a memory state $\alpha_{t} = \{1;01\}$, the circuit would return $p_{t} =(111)$. 
}
\label{fig:single_conv_in_out}
\end{figure}

Similarly 
\begin{align}
H(D) &\equiv H^{(1)}+D H^{(2)}+...+D^{\nu_h-1} H^{(\nu_h)},\nonumber\\
&\equiv \left[\begin{array}{cccc}
h_{11}(D) & h_{12}(D) & \hdots & h_{1n}(D)\\
\vdots &        &        & \vdots\\
h_{z1}(D) & h_{z2}(D) & \hdots & h_{zn}(D)
\end{array} \right]_{{n_z}\times n},\nonumber
\end{align}
where $h_{ij}(D) = \sum\limits_{q = 1}^{\nu_h} D^{q-1}H^{(q)}_{ij}
$.

In this delay notation, \cref{eq:commutation_conditions} becomes
\begin{align}
\hspace{-3mm}D^{u}\!\left[\begin{array}{c}G_Z\transpose(D)\\H_{Z}(D)\\ \text{ISF}_Z(D) \end{array} \right]\!\!\left[\!\begin{array}{lll}\!G_X(\widetilde{D})\!& \!\text{ISF}\transpose_X(\widetilde{D}) & \!H_X\transpose(\widetilde{D})\! \end{array} \!\right]\!\widetilde{D}^{v} &\!=\! \mathbb{I}_{n\times n }\delta_{uv},
\label{eq:commutation_conditionsDelay}
\end{align}

Since convolutional codes are partitioned into frames, we write $\vec p=(\vec p_1,\vec p_2,..., \vec p_t,...,\vec p_\tau)$, where $\vec p_j\in \mathbb{Z}_2^n$. From \cref{pvec} we have $\vec{p} =\sum_i^{\tau}\vec{p}_t D^{i-1}$, and
\begin{align}
\vec{p}_t =& \sum\limits_{i = 1}^{k}\sum\limits_{j = 1}^{
\nu_g+1}l^{(g)}_{i,t-j+1}  G\transpose_i(D) \widetilde D^{j-1} \nonumber\\&
+\sum\limits_{i = 1}^{{n_z}}\sum\limits_{j = 1}^{\nu_h+1}l^{(h)}_{i,t-j+1}  H_i(D)\widetilde D^{j-1},\nonumber\\
    \equiv& U_p (\alpha_t, \vec{l}_t ),\label{UP}
\end{align}
where $\vec{l}_t\equiv\{\vec{l}^{(g)}_t ; \vec{l}^{(h)}_t\} \equiv\{l_{1,t}^{(g)},...,l_{k,t}^{(g)}; l_{1,t}^{(h)},...,l_{n_z,t}^{(h)}\}$ and 
\begin{equation}
\alpha_t\equiv \{ \vec{l}^{(g)}_{t-1},\vec{l}^{(g)}_{t-2},...,\vec{l}^{(g)}_{t-\nu_g} ;\vec{l}^{(h)}_{t-1},\vec{l}^{(h)}_{t-2},...,\vec{l}^{(h)}_{t-\nu_h} \}.\label{alpha}
\end{equation}
The \emph{memory state} $\alpha_t$ is a list of length $\nu_g+\nu_h$ that records values of $\vec l_q$ for $q=t-\nu_{g,h}+1$ up to $q=t-1$. 
It becomes useful when we discuss a SISO decoder in Section \ref{sec:trellis_siso_algorithm}, which optimises the choice of $l$'s.

The purpose of these manipulations is that $\vec{p}_t$ depends only on prior  $\vec l_t$'s  going back  a number of frames depending on the memory length $\nu_{g,h}$. Importantly, it does not grow with $\tau$, so that the search-space for optimising over $l$'s does not grow exponentially with the size of the code. 

From \cref{alpha} it is simple to see the memory state $\alpha_{t+1}$ depends on  $\vec l_t$, and components of $\alpha_t$. As a result, there are consistency conditions that relate $\alpha_{t+1}$ to $\alpha_{t}$.  These conditions are represented by a \emph{trellis}, which depicts all valid \emph{transitions} from $\alpha_{t}$ to $\alpha_{t+1}$.

\begin{figure}[t]

\includegraphics{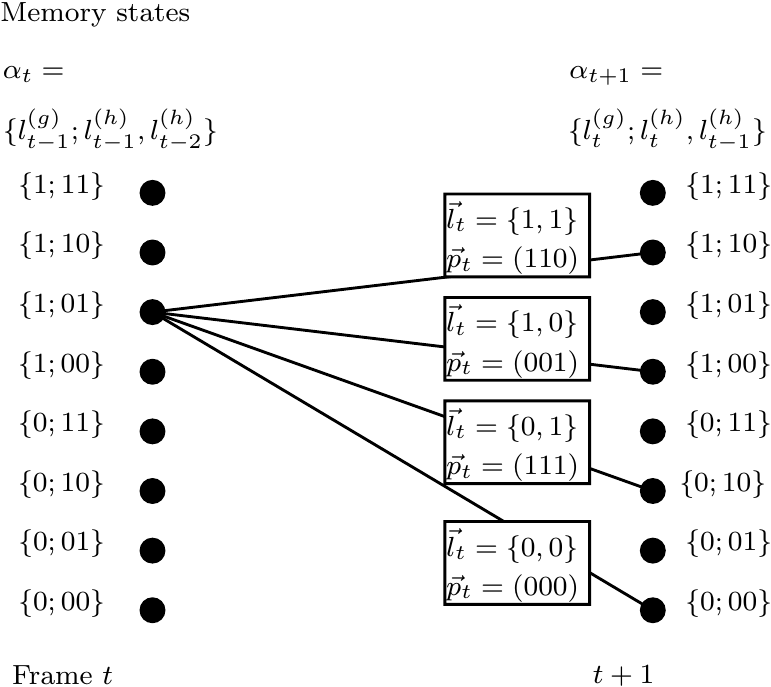}
\caption{The structure of transitions between memory states corresponding to the example illustrated in \fig{fig:single_conv_in_out} using the convolutional code defined in\cref{eq:example_conv}. The recovery operation is determined by  $\vec{p}_t = U_p(a_t,\vec{l}_t) = \{111\}$. Enumerating over all possible logical inputs, $\vec{l}_t$, starting memory states, $a_t$, and frames, $t$, a complete trellis is constructed, shown in \fig{fig:conv_trellis}.  }
\label{fig:single_transition}
\end{figure}

\begin{figure}[t]

\includegraphics{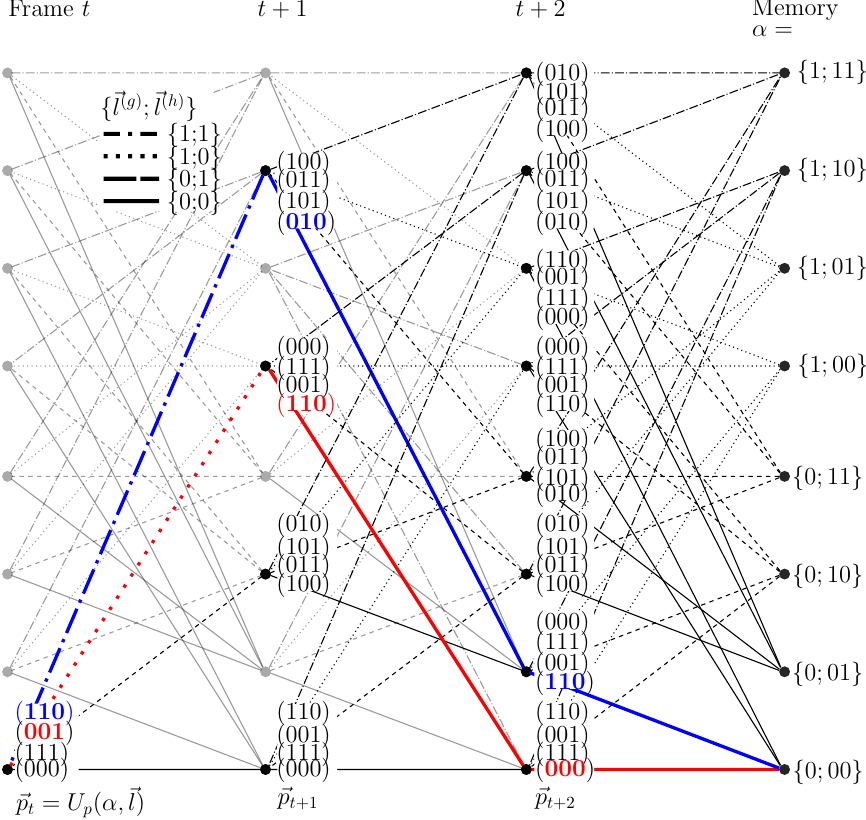}
\caption{The decoding trellis for the convolutional code in\cref{eq:example_conv}. Each transition corresponds to a different product of $\hat{g}$ and $\hat{h}$ at a given time-step. The memory state records the values of previous transitions. Each path corresponds to a certain product of $\hat{g}$ and $\hat{h}$ terms. The blue and red paths are the most likely paths for the syndromes in \cref{eq:blue_path} and \cref{eq:red_path} respectively.}
\label{fig:conv_trellis}
\end{figure}

We now illustrate this construction for the example of a $\tfrac{k}{n}=\tfrac{1}{3}$ convolutional code. \fig{fig:single_conv_in_out} shows a (classical) circuit diagram relating  $\vec{p}_t$ to $\vec l_t$, for the self-dual rate convolutional code defined by
\begin{align}
G_Z\transpose(D) &= \left[\begin{array}{C{20mm}C{15mm}C{10mm}} $D$ & $D$ & $1$ \end{array} \right],\nonumber\\
H_Z (D)  &= \left[ \begin{array}{C{20mm}C{15mm}C{10mm}} $1+D+D^2$& $1+D^2$ & $1$ \end{array} \right],\label{eq:example_conv} \\
\text{ISF}_Z(D) &= \left[ \begin{array}{C{20mm}C{15mm}C{10mm}} $D$& $D$ & $0$ \end{array}  \right].\nonumber
\end{align}
Reading off the highest powers of $D$ in $G_Z\transpose(D)$ and $H_Z(D)$ respectively, we see that  $\nu_g = 1$ and $\nu_h = 2$, so that the memory state \mbox{$\alpha_t=\{l^{(g)}_{t-1};l^{(h)}_{t-1},l^{(h)}_{t-2}\}$} is a list with three entries. In the equivalent circuit depicted in \fig{fig:single_conv_in_out}, the memory is represented by black boxes.

\fig{fig:single_transition} illustrates allowed transitions from state $\alpha_t=\{1;01\}$ to a new state $\alpha_{t+1}$ at frame $t$ for the convolutional code defined in \cref{eq:example_conv}. Given that this code example is sufficiently simple, we show all possible choices for $\vec l_t$ for this transition, i.e.\  $\vec{l}_t = \{0;0\}$, $\vec{l}_t = \{0;1\}$, $\vec{l}_t = \{1;0\}$  and $\vec{l}_t = \{1;1\}$.  We also show the corresponding codeword block frame $\vec p_t$, respectively  $(000)$, $(111)$, $(001)$ and $(110)$  for each possible choice of $\vec{l}_t$.  Expanding \fig{fig:single_transition} over several frames yields the trellis, shown in  \fig{fig:conv_trellis}.  

With the foregoing machinery in place, error correction works as follows. A set of errors, $\vec{\varepsilon}$, produces a syndrome $\vec S$. From $\vec S$ we use the ISF to calculate $\vec{e}^{\,0}$.   Any  path, $\vec p=(\vec p_1,\vec p_2,..., \vec p_t,...,\vec p_\tau)$, through the trellis that begins in the trivial state $\vec{p}_1=(0...0;0...0)$ and ends  in the state $\vec{p}_\tau=(0...0;0...0)$ state corresponds to a valid decoding. 
The path $\vec{p}_\text{min}$ which minimises the Hamming weight of \mbox{$\vec{p}_\text{min} + \vec{e}^{\,0}$} is the most likely decoding solution.  If no detectable errors occur so that $\vec{e}^{\,0}$ is trivial (i.e.\ if $\vec{e}^{\,0}=\{00...\}$), then the lowest weight path through the trellis has $\vec{p}_\text{min}=(000...)$, and $\text{wt}(\vec{e}^{\,0}+\vec{p}_\text{min})=0$.  For nontrivial $\vec{e}^{\,0}$, finding this path is the core of the decoding problem.

Given a trellis, such as in \fig{fig:conv_trellis}, and priors for the qubit error marginals, a   trellis decoding algorithm will return a suitable choice for $\vec{p}_\text{min}$.  Several algorithms exist  to identify (optimally or suboptimally) the most likely trellis path. These include the Viterbi algorithm\cite{viterbi_1967} the Bahl, Cocke, Jelinek, Raviv (BCJR) algorithm\cite{bahl_1974} and the Benedetto algorithm\cite{benedetto_1996}.

The last of these algorithms is a maximum \emph{a posteriori} (MAP) SISO decoder which is capable of dealing with parallel transitions, i.e.\ those for which several transitions between  states $\alpha_t$ and $\alpha_{t+1}$ exist. Single layer foliated codes necessarily  have parallel transitions in their trellis descriptions, as will be shown in section~\ref{sec:fol_conv_trellises}. For this reason we will use the Benedetto algorithm as a SISO decoder for convolutional codes.  This is discussed in detail in Section \ref{sec:trellis_siso_algorithm}.

 \subsubsection*{Examples}
 
 Before concluding this section, we give two illustrative examples using the trellis shown in \fig{fig:conv_trellis}. Because the examples are sufficiently simple, we find the  lowest weight path through the trellis by inspection. In practice we use a modified version of Benedetto's trellis algorithm \cite{benedetto_1996} to determine the most likely path through the trellis (see \S \ref{sec:trellis_siso_algorithm}).
 
 In the first example the error pattern, $\vec\varepsilon$, is a single error occurring on the first bit in frame $t+1$, as indicated in  the following equation:
 \begin{align}
\begin{array}{r c ccccc c}
  \textit{frame:}         &\hdots &t-2&t-1&t  &t+1&t+2&\hdots\\
\vec\varepsilon   =&\hdots &000&000&000&100&000&\hdots\\           
\vec S        =&\hdots & 0 & 1 & 1 & 1 & 0 &\hdots\\
\vec{e}^{\,0} =&\hdots &000&000&110&110&110&\hdots\\
\vec{p}_B =&\hdots &000&000&110&010&110&\hdots\\
\vec{e}_{\text{min}}  =& &\multicolumn{5}{l}{\vec{e}^{\,0} + \vec{p}_B} & \\
 =&\hdots &000&000&000&100&000&\hdots
\end{array},
\label{eq:blue_path}
\end{align}
 Given, $\vec\varepsilon$ we find $\vec S$ and $\vec{e}^{\,0}$, also shown in  \cref{eq:blue_path} (for this and the following example, we work through the calculation of $\vec S$ and $\vec{e}^{\,0}$ from $\vec \varepsilon$ in Appendix \ref{eScalc}). Using $\vec{e}^{\,0}$, we find (by inspection for this example)  the  path $\vec{p}_B$ through the trellis that minimises $\text{wt}(\vec{e}^{\,0}+\vec{p}_B)\equiv\text{wt}(\vec{e}_{\text{min}})=1$.    This path is shown in blue in \fig{fig:conv_trellis}; the blue highlighted triples of binary values in the figure correspond, frame-by-frame, the triples in $\vec{p}_B$. This path determines the optimal error correction procedure, which is to apply the correction operator $Z^{\otimes \vec{e}_{\text{min}}}$.  \tmsB{In this example, $\vec{e}_{\text{min}}=\vec \varepsilon$, so the error correction would succeed since $\vec{e}_{\text{min}}$ and $\vec \varepsilon$ are logically equivalent,  i.e.\ the correction $\vec{e}_{\text{min}}$  would return the system to the correct codeword.} 

In the second example the error pattern, $\vec\varepsilon$, consists of two errors occurring on the first and second qubits in frame $t+1$. The most likely path through the trellis, $\vec{p}_R$, is shown in red in \fig{fig:conv_trellis}. We have
\begin{align}
\begin{array}{r c ccccc c}
      \textit{frame:}     &\hdots &t-2&t-1&t  &t+1&t+2&\hdots\\
\vec\varepsilon  =&\hdots &000&000&000&110&000&\hdots\\           
\vec S         =&\hdots & 0 & 0 & 1 & 0 & 0 &\hdots\\
\vec{e}^{\,0} =&\hdots &000&000&000&110&000&\hdots\\
\vec{p}_R =&\hdots &000&000&001&110&000&\hdots\\
\vec{e}_{\text{min}}  =& &\multicolumn{5}{l}{\vec{e}^{\,0} + \vec{p}_R} & \\
 =&\hdots &000&000&001&000&000&\hdots
\end{array}.
\label{eq:red_path}
\end{align}
\tmsB{Again, the red highlighted triples of binary values in  \fig{fig:conv_trellis} correspond, frame-by-frame, the triples in $\vec{p}_R$ in \cref{eq:red_path}.} The lowest weight recovery operation is a single error on the third qubit in frame $t$. In this example, the product of the recovery operation and physical errors is $ \vec{e}_\text{min} + \vec\varepsilon=...001\,110...$ is a nontrivial codeword of $G_Z\transpose$ in \cref{eq:example_conv}.  That is the decoder fails in this example. 

This second example is illustrative: because this is $d=3$ code, adjacent errors are not expected to be corrected.  However, if errors are sufficiently far apart (determined by the code memory length), then they behave as if they were independent, and so the convolutional code can decode many more errors than $d/2$ if they are sparsely distributed.

\section{Foliated Convolutional Decoding}
\label{sec:fol_conv_trellises}

\begin{figure}[t]
\includegraphics[width=\columnwidth]{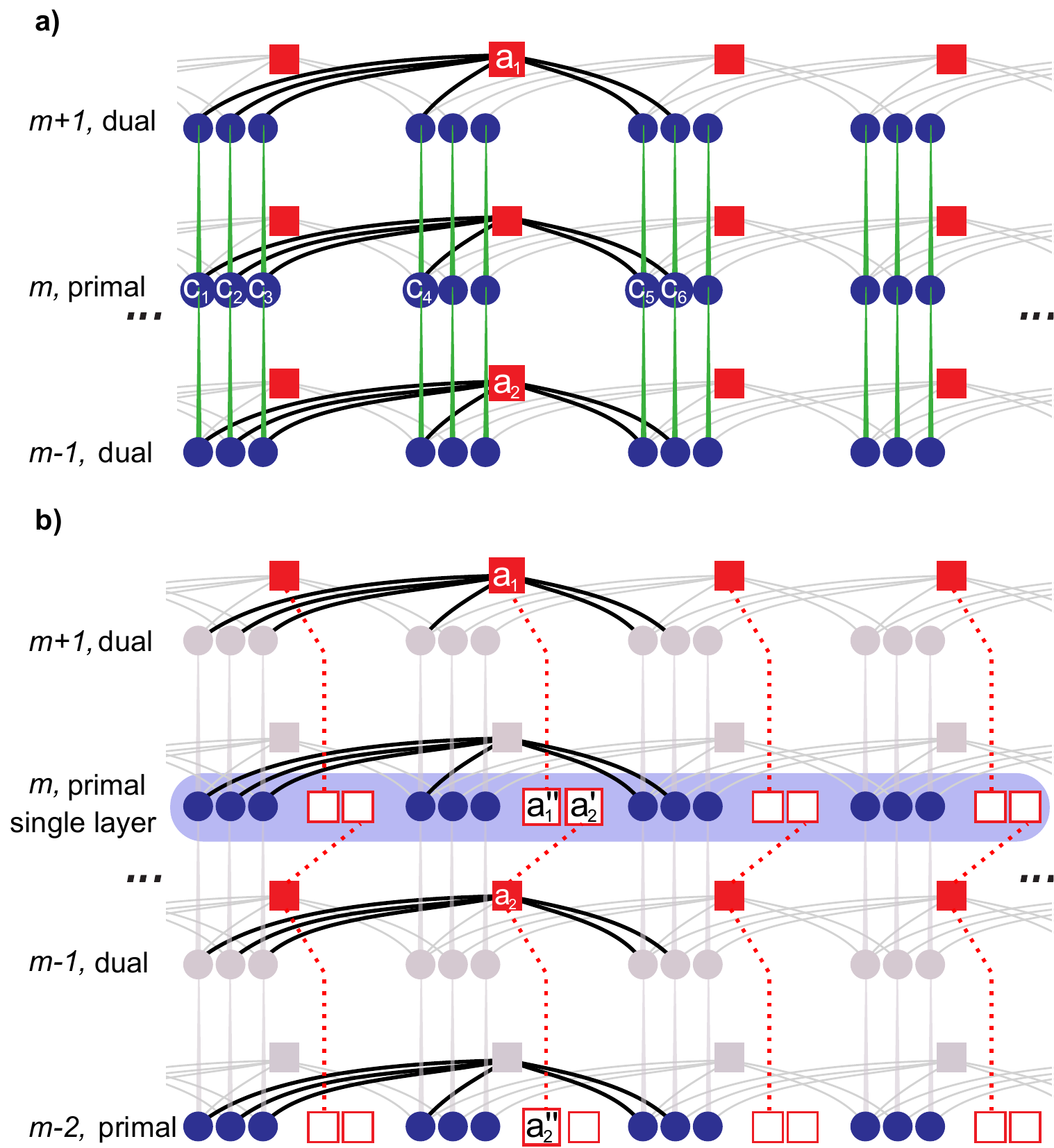}
\caption{{\bf a)} The foliated rate $\frac{1}{3}$ convolutional code defined by equations~\ref{eq:example_conv}. The product of cluster stabilisers centred on the numbered qubits generates a parity check operator \mbox{$\hat P=K_{c_1}K_{c_2}\hdots K_{a_2} = X_{c_1}X_{c_2}\hdots X_{a_2}$}. All other parity check operators are translations of this seed.  {\bf b)} The corresponding primal lattice. Dual elements have been grayed out. Because there is only $n_x=1$ $X$-like stabliser per frame, we introduce $2n_x=2$ virtual qubits, labelled $a'_i$  and $a''_i$, corresponding to ancilla qubits $a_{1,2}$ on neighbouring sheets. This results in a rate $\frac{k}{n+2n_x}=\frac{1}{5}$ for sheet $m$ of the foliated structure. The marginals on virtual qubit $a_1''$, computed within sheet $m$, and virtual qubit $a_1'$ in sheet $m+2$ (which ultimately refer to the same physical qubit $a_1$) are exchanged iteratively, so that the foliated decoder converges. 
}
\label{fig:marginal_exchange}
\end{figure}

In this section we build on the trellis construction in section~\ref{sec:convolutional_trellis_construction} to operate on independent sheets of a foliated convolutional code, accounting for additional ancilla. 
Decoding can be performed using the algorithm in section~\ref{sec:trellis_siso_algorithm}. This represents a sub-routine in the full foliated decoding algorithm which exchanges marginals between sheets. This is reviewed in section~\ref{sec:marginal_exchanges}.

\subsection{Trellises for Single Decoding Sheets}\
\label{sec:trellises_single_layer}

Consider a foliated code based on a rate $r={k}/{n}$ CSS convolutional code with ${n_z}=|\mathcal{S}_Z|/\tau$ $Z$-like stabilisers and \mbox{${n_x}=|\mathcal{S}_X|/\tau$} $X$-like stabilisers per frame, so that there are $n = k+n_x+n_z$ physical qubits per frame. Foliated parity check operators are defined in \cref{eq:foliatedparitycheck}. \fig{fig:marginal_exchange}a shows an example of a foliated rate $1/3$ convolutional code, and a specific parity check operator, \tmsB{$\hat P=X_{a_1}X_{c_1}...X_{c_6}X_{a_2}$, is indicated by the labelled vertices ${a_1},{c_1},...,{c_6}$ and ${a_2}$, where the ancilla qubits are on the adjacent code sheets $m\pm1$}.  Measuring all such operators associated to code sheet $m$ yields the syndrome $\vec S_m$ for that sheet.  

For the purposes of decoding, we introduce \emph{decoding sheets}. A decoding sheet $m$ is identified with the corresponding code sheet $m$:  it refers to the same code qubits, but includes virtual ancilla qubits associated to the neighbouring sheets. 

 The shaded elipse in \fig{fig:marginal_exchange}b illustrates the formation of a single decoding sheet for a  foliated convolutional code; this decoding sheet, $m$, is identified with the corresponding code sheet, $m$, in \fig{fig:marginal_exchange}a. 
 Independent decoding of each decoding sheet can be performed by creating a virtual code associated with a single sheet of the foliated structure.  This code accounts for the ancillary qubits $a_1$  and $a_2$ in adjacent sheets by introducing virtual ancilla  $a_1''$  and $a_2'$. There are $2n_x$ virtual ancilla associated to the ancilla on the adjacent code sheets, so the virtual code has rate $\frac{k}{n+2n_x}$.

 We note that a physical ancilla qubit $a_{j,m}$ is represented virtually in two different decoding sheets $m\pm1$ (as $a_{j,m}'$ and $a_{j,m}''$). This yields a consistency condition on ancilla marginals between neighbouring sheets, which we discuss later.

Within a decoding sheet (see \fig{fig:marginal_exchange}b), we begin by finding an initial recovery operation, $\hat{E}^0$, which satisfies a received syndrome, $\vec S$. As with the previous section this can be achieved by using ISFs. Setting the final $2n_x$ qubits in a frame to correspond to virtual ancilla qubits, the seed generator is given by
\begin{align}
\overline G\transpose_Z(D) = \left[\begin{array}{C{1.5cm}|C{1.1cm}C{1.1cm}}$G^{T}_Z(D)$&$\mathbf{0}_{k\times n_x }$&$\mathbf{0}_{k\times n_x }$ \end{array}\right],
\end{align}
where $G^{T}_Z$ refers to the generator matrix of the base convolutional code, as exemplified in \cref{eq:example_conv}. Similarly the seed parity check operators and ISF are given by
\begin{align*}
\overline P (D)= \left[\begin{array}{C{1.5cm}|C{1.2cm}C{1.2cm}}$H_X(D)$&$\mathbb{I}_{n_x\times n_x}$&$\mathbb{I}_{n_x\times n_x}$\end{array} \right],\\
\overline \ISF_Z (D)= \left[\begin{array}{C{1.5cm}|C{1.2cm}C{1.2cm}}$\ISF_Z(D)$&$\mathbf{0}_{n_x\times n_x}$&$ \mathbf{0}_{n_x\times n_x}$\end{array} \right],\\
\overline H_Z(D) = \left[ \begin{array}{C{1.5cm}|C{1.2cm}C{1.2cm}}$H_Z(D)$&$\mathbf{0}_{n_z\times n_x}$ &$ \mathbf{0}_{n_z\times n_x}$\end{array} \right].
\end{align*}
Additional pairs of gauge operators are generated from the degrees of freedom introduced by the extra ancilla qubits. The seed generators, stabilisers, ISFs and gauges $J_Z$ and $J_X$ satisfy the orthogonality relations
\begin{equation}
D^i\! \!\left[\begin{array}{r} \overline G_Z\transpose(D) \\
\overline  H_Z(D) \\
\! \overline{\text{ISF}}_Z(D) \\
  \overline J_Z(D) \end{array} \right]\!\!
 \left[ \begin{array}{llll}\overline G_X\transpose(\widetilde{D}) & \overline {\text{ISF}}_X(\widetilde{D}) & \overline P(\widetilde{D}) & \overline J_X(\widetilde{D}) \end{array} \right] \!= \!\mathbb{I}\delta_{i0},
\end{equation}
which implicitly defines $\overline J_{X,Z}$. A valid choice of $\overline J_X$ is 
\begin{align}
\overline J_X = \left[\begin{array}{c|cc}\mathbf{0}_{n_x\times n}&\mathbb{I}_{n_x\times n_x} &\mathbf{0}_{n_x\times n_x}\\
\mathbf{0}_{n_x\times n} &\mathbf{0}_{n_x\times n_x}&\mathbb{I}_{n_x\times n_x}
\end{array} \right],\end{align}
which orthogonal to $G_Z$, $H_Z$ and $\ISF_Z$.  

Generally, $\overline J_Z$ depends on the details of the code. 
Each $Z^{\otimes \vec{j}} \in J_Z$ is a set of operators which commute with $G_X$, $P_X$ and $\ISF_X$, but anti-commute with a single $X^{\otimes a} \in J_X$. 

As an example, foliating the code in \cref{eq:example_conv}, we have 
\begin{align}
\overline J_Z(D) &= \left[\begin{array}{ccc|cc}
1+D&1&1&D&0 \\ 0&0&0&1&1 
\end{array} \right].
\label{eq:C3_J}
\end{align}
Because $J_Z$ commutes with the stabilisers, they correspond to undetectable  error patterns within the sheet.  As an aside, using the Raussendorf lattice as an example, these would correspond to error chains that pass through a sheet in the `time'-like direction of foliation: they leave no syndrome data within the sheet (but would be detected by other sheets in the 3D lattice).

The set of valid recovery operations is given by
\begin{align}
 \mathcal{E} =\{Z^{\otimes \vec{e}^{\,0} + \vec h + \vec g + \vec J}|&
 \vec h\!\in \!\text{RowSpace}(\mathbf{H}_Z\!), \nonumber\\&
 \vec g\in \text{RowSpace}(\mathbf{G}\transpose_Z\!), \nonumber\\&
 \vec{J} \in \text{RowSpace}(\mathbf{J}_Z)\}.\nonumber
\end{align}
where 
\begin{align}
\mathbf{J}_Z \!\!=\! \left[ \begin{array} {c cc ccc c}
\hdots& J^{(1)}        &\hdots & J^{(\nu_j)} &\hdots  &      &\\
& \hdots     & J^{(1)}    &\hdots   & J^{(\nu_j)}&\hdots&\\
  &         & \hdots    & J^{(1)}       & \hdots &J^{(\nu_j)} &\hdots
\end{array}\right]_{2\tau\times n\tau},
\end{align}
by analogy with \cref{eq:fullgenerator}.  In the example in  \cref{eq:C3_J}, 
\begin{equation}
 J^{(1)} = \left[\begin{array}{ccc|cc}
1&1&1&0&0 \\ 0&0&0&1&1 
\end{array} \right],\text{ and }J^{(2)} = \left[\begin{array}{ccc|cc}
1&0&0&1&0 \\ 0&0&0&0&0 
\end{array} \right].
\end{equation}

 Given $\mathbf{H}_Z$, $\mathbf{G}\transpose_Z$ and $\mathbf{J}_Z$, a trellis may be constructed, closely mirroring  the trellis construction in section \ref{sec:convolutional_trellis_construction}.  The main difference is that the decoding trellis must be modified to account for $\overline J_Z$ terms, which represent the virtual ancilla in each sheet.  This give rise to larger memories and more allowed transitions. The $\vec{l}$ terms are modified to include $l^{(J)}_t$ components.  

Once the trellis is formed, a soft-input hard-output decoder will take as input marginal error probabilities on the qubits, and return $\vec h$, $\vec g$ and  $\vec J$, which correspond to a specific, maximum likelihood error pattern consistent with the syndrome data on sheet $m$.  More suited to our goal is a soft-input soft-output (SISO) decoder which returns  updated \emph{a posteriori} probabilities for qubit error marginals.  We discuss  such a decoder in  Section \ref{sec:trellis_siso_algorithm}.

\subsection{Ancilla Marginal Exchange for Consistency Between Decoding Sheets}
\label{sec:marginal_exchanges}

When decoding is performed on each   convolutional code sheet, a given ancilla qubit, $a_{k,m}$, is associated to corresponding virtual ancilla, $a_{k,m}''$ and $a_{k,m}'$,  from each of the neighbouring decoding sheets $m+1$ and $m-1$. For consistency, we require the SISO marginal \emph{a posteriori} error probabilities on $a_{k,m}'$ and $a_{k,m}''$ to match, i.e.\  $P(a_{k,m}'')=P(a_{k,m}')$. 

Independent SISO decoding of adjacent decoding sheets does not automatically respect this constraint, so we perform sequential rounds of independent decoding, and iteratively exchange marginals, $P(a_{k,m}'')\leftrightarrow P(a_{k,m}')$ between each round, until the marginals are satisfactorily converged.

This approach is similar to belief propagation on a modified Tanner graph, with a message passing schedule which gives priority to intra-sheet messages over that of inter-sheet messages~\cite{tanner_1981,mackay_1997,mackay_2004,pearl_1985}. It is unlikely that the process we have proposed is optimal, but good numerical decoding results are still possible as shown in section~\ref{sec:numerical_results} for turbo codes.

In practical settings it may be desirable to terminate the iterative message passing  after a fixed number of rounds.  In this case, the marginals on the ancilla from adjacent sheet decoders may not have converged.  One heuristic resolution to this is to average the inconsistent marginals on each ancilla, $P(a_{k,m}):=(P(a_{k,m}'')+ P(a_{k,m}'))/2$, and then make a hard decoding choice, by choosing the most likely error configuration at each ancilla.  A final decoding round is performed on each sheet.
   This ensures that the decoding falls within the codespace.

\section{Trellis SISO Algorithm}
\label{sec:trellis_siso_algorithm}

So far, we have concentrated on constructing trellises for convolutional codes.  As already mentioned, we then use the trellis to  find optimal error configurations (i.e.\ a hard decoding) or assign error marginals  (i.e.\ soft decoding) that are consistent with the syndrome data.  

In this section we present a modified MAP SISO trellis decoder, following the approach of \citet{benedetto_1996} on classical convolutional codes. The modified algorithm compares codewords and stabilisers against a trial  error pattern, $\vec{e}^{\,0}$, determined from the syndrome data, and calculates marginal error probabilities consistent with it. 

The algorithm consists of three stages: (1) a forward pass, which passes through the trellis from left to right, (2) a backward pass, which passes through the trellis from right to left, and (3) a local update which determines marginals using information from forward and backward passes.

In the rest of this section, we detail the core of the SISO algorithm. As in prior sections, the trellis memory states at frame $t$ is $\alpha_t$. The initial error pattern $\vec{e}^{\,0}_t$ is calulated from the syndrome, $\vec S$, and the ISF. The physical operations associated with a transition through the trellis are denoted $\vec p_t$.

\subsection{Forward Pass Algorithm}

The forward pass algorithm assigns a probability for each memory state,  $A(\alpha_t|\vec S_{i \leq t})$, as we pass through  the trellis starting from the first frame, $t=1$, (i.e.\  the leftmost frame in the presentation of \fig{fig:conv_trellis}) and traversing to $t=\tau$ (right).  Here $\vec S_{i \leq t}$ is a shorthand notation which indicates that only syndrome information up to frame $t$ is used to calculate likelihoods.  At each frame, there are $2^{(k\nu_g+n_z\nu_h)}$
memory states to store. Note that for decoding sheets in the foliated construction a memory term $\nu_J$ must be incorporated for the $J_Z$ guages.

We assign initial probabilities   \mbox{$A(\alpha_1 = \vec{0}) = 1$} and \mbox{$A(\alpha_1 \neq \vec{0}) = 0$ } to memory states at frame $t=1$.  
The forward pass algorithm then computes probabilities for subsequent memory states as 
\begin{align}
A(\alpha_{t+1}|\vec S_{i \leq t} ) &= \sum\limits_{\vec{l}_t}A(\alpha_{t}|\vec S_{i \leq t-1})\text{Pr}(\vec{p}_t +  \vec{e}^{\,0}_t)\text{Pr}(\vec{l}_t).
\label{eq:forward_pass}
\end{align}
where $\vec{p}_t = U_p(\alpha_{t},\vec{l}_t)$ is given by \cref{UP}, and we recall from \cref{alpha} that $\alpha_{t}$ on the RHS of \cref{eq:forward_pass} is determined by   $\alpha_{t+1}$ and $\vec l_t$.  $\text{Pr}(\vec{p}_t +  \vec{e}^{\,0}_t)$ is the \emph{a priori} probability of the error pattern $\vec{p}_t +  \vec{e}^{\,0}_t$, which  depends on the details of the prior error mode; for our purposes this is just an i.i.d.\ error process on each of the physical qubits, so that $\text{Pr}(\vec{p}_t +  \vec{e}^{\,0}_t)$ is given by the binomial formula.  $\text{Pr}(\vec{l}_t)$ is the \emph{a priori} probability of undetectable  error processes (generated by \cref{pvec}).  For convolutional codes, we take  $\text{Pr}(\vec{l}_t)$ to be a constant (so that it factors out of $A$),  however it will become important when we consider turbo codes, in which physical errors in the inner code affect logical  priors in the outer code.

\subsection{Backward Pass Algorithm}

The backward pass algorithm is identical to the forward pass, but working now from the last frame, $t=\tau$ back to the first, $t=1$.  The update rule is given similarly,
\begin{align}
B(\alpha_t|\vec S_{i \geq t1}) &= \sum\limits_{\substack{\vec{l}_t}}
B(\alpha_{t+1}|\vec S_{i \geq t+1}) \text{Pr}(\vec{p}_t +  \vec{e}^{\,0}_t)\text{Pr}(\vec{l}_{t}).
\end{align}
The initial conditions are set as $B(\alpha_\tau= \vec{0}) = 1$ and $B(\alpha_\tau \neq \vec{0}) = 0$, where $\tau$ is the final frame.

\subsection{Local Update Algorithm}

The local update calculates the likelihoods of physical error patterns, $ \vec  e_{p,t}=\vec{p}_t+\vec{e}^{\,0}_t\in\mathbb{Z}_2^n$, at frame $t$, given a valid error configuration $\vec{e}^{\,0}_t$ (which is itself derived directly from the syndrome data). That is, we calculate the marginals  $P(\vec  e_{p,t}|\vec S)$,  and the marginals over logical  bits $P( \vec e_{l,t}|\vec S)$, where $\vec e_l\in\mathbb{Z}_2^{k+n_z}$ is a specification of the logical states at frame $t$.  These marginals depend  in turn on the  marginal beliefs of memory states computed in  the forward, $A$, and backward, $B$, pass algorithms. As inputs the decoder uses an initial prior  distribution, $\text{Pr}(\vec{e}_{p,t})$, for the error configuration $\vec e_{p,t}$, and the logical error patterns, $\text{Pr}(\vec{l} = \vec e_{l,t})$, as well as the syndrome, and computes the marginals
\begin{align}
P(\vec e_{p,t}|\vec{e}^{\,0}_t ) \!= \mathcal{N}_{p}\hspace{-6mm} \sum\limits_{\substack{\vec{l}_t,\alpha_t:\\\vec{p}_t = U_p(\alpha_t,\vec{l}_t)  
 }}\hspace{-5mm} 
A(\alpha_t)B(\alpha_{t+1})\text{Pr}(\vec{p}_t+\vec{e}^{\,0}_t )\text{Pr}(\vec{l}_t), 
\end{align}
\begin{align}
P(\vec e_{l,t}|\vec{e}^{\,0}_t ) \!= \mathcal{N}_{l}\hspace{-6mm} \sum\limits_{\substack{\alpha_t:\\\vec{p}_t = U_p(\alpha_t,e_{l,t})
}}\hspace{-5mm}
A(\alpha_t)B(\alpha_{t+1})\text{Pr}(\vec{p}_t+\vec{e}^{\,0}_t)\text{Pr}(\vec  e_{l,t}),
\label{eq:local_update2}
\end{align}
where $\mathcal{N}_{p}$ and $\mathcal{N}_{l}$ are normalization constants chosen to ensure that $\sum_{\vec e_{p,t}}P(\vec e_p|\vec{e}^{\,0}_t )=1$ and  $\sum_{\vec e_{l,t}}P(\vec e_l |\vec{e}^{\,0}_t )=1$. 
 Again, we recall from \cref{alpha} that $\alpha_{t}$ on the RHS of \cref{eq:forward_pass} is determined by   $\alpha_{t+1}$ and $\vec l_t$.

Equations \ref{eq:forward_pass} to \ref{eq:local_update2} are the ingredients for the \emph{sum-product} belief propagation algorithm: the Forward and Backward Pass algorithms each run independently, and then the Local Update calculates marginals for each of the $2^n$ possible error configuration over each of the $\tau$ frames. Storing this information requires memory $\sim \tau 2^n$.

As an aside, the \emph{max-sum} algorithm, which has some practical performance benefits,  approximates the \emph{sum-product} algorithm, by summing over logs of marginals ~\cite{bahl_1974}.

\section{Foliated Turbo Codes}
\label{sec:numerical_results}

Our main motivation for studying turbo codes is  to demonstrate the foliated construction and BP decoder in an extensible, finite-rate code family.  Practically, these and other finite-rate codes may have  applications in \tmsA{fault-tolerant  quantum repeaters networks  \cite{li_2010,duan_2001,1367-2630-7-1-194,1367-2630-12-9-093032},  where  local nodes  create optimal clusterised  codes 
to reduce resource overheads or error tolerance \cite{satoh_2016}}, however we do not address these applications here.

\subsection{Turbo Code Construction}
\label{turboconstruction}

A turbo codes is essentially a concatenation of two convolutional codes, albeit with an \emph{interleaver} between them.  When convolutional codes fail, they tend to produce bursts of  errors on logical bits.  Turbo codes address this by concatenating encoded (qu)bits from the \emph{inner} convolutional code into widely separated logical (qu)bits in the \emph{outer} code.  The interleaver is simply a  permutation, $\Pi$, on the inner logical qubits, and serves to transform a local burst of errors from the inner decoder into widely dispersed (and thus approximately independent) errors that the outer decoder is likely to correct.  

For the numerical results we present in this section, we choose $\Pi$ to be a completely random permutation on the inner code.  This is a conventional choice for benchmarking turbo codes, however a completely random permutation leads to highly delocalised encodings.  This may be undesirable in the quantum setting, and the optimal choice of interleavers \tmsB{was discussed in} \cite{sadjadpour_2001,vafi_2005}.  In the context of constructing clusterised codes, the interleaver choice will affect the weight of stabilisers.  In section \ref{sec:turbo_code_arch} we return to this issue, and show that by choosing a structured interleaver, we reduce the weight of stabilisers substantially. This reduces the weight of correlated errors that build up during the systematic construction of the  cluster state resource.

Turbo codes are generated using underlying convolutional codes. We use two different convolutional codes, one with $d=3$, which we refer to as the C3 family of codes, and one with $d=5$, which we refer to as the C5 family\cite{forney_2007}. When embedded as clusterized codes the distance of these codes are reduced to 2 and 3 respectively. This represents the effective code distance, $d_{\text{eff}}$, of a single clusterised code sheet, which forms part of the larger foliated structure. While the effective distances are diminished for codes acting within a single sheet, the distance of the foliated convolutional codes remain 3 and 5 respectively. This is because some error patterns which are undetectable within a single sheet decoder are detectable by neighbouring layers, as discussed in section~\ref{sec:fol_conv_trellises}.

At the end of this section, we present numerical results about the decoding performance of two families of foliated turbo codes, T9 and T25 codes. T9 codes are generated from the concatenation and interleaving of two C3 codes; similarly the T25 code is formed from the concatenation and interleaving of two C5 codes. The distances of these  turbo codes are $d_{T9}=9$ and $d_{T25}=25$ respectively.

\subsection{Turbo Code Decoders}
\label{sec:turbo_decoder}

\begin{figure}
\includegraphics{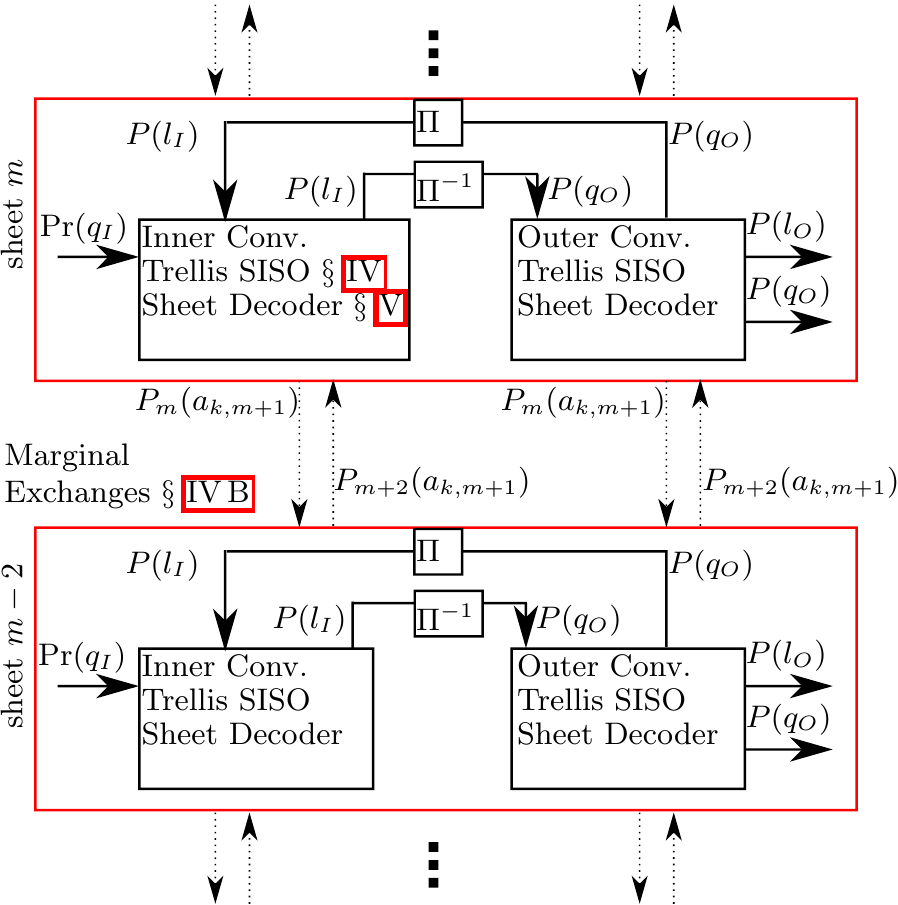}
\caption{The decoding process for foliated turbo codes. An inner sheet convolutional decoder is run for each sheet in the foliated code. Ancillas marginals are exchanged between neighbouring primal (dual) layers, ie next-nearest neighbours, shown as dotted arrows. The process is then iterated. The logical marginals, $P(l_I)$ are then de-interleaved ($\Pi^{-1}$) and used by the outer sheet convolutional decoders. Ancilla marginals are exchanged, and as before the process is iterated. The outer sheet decoders pass physical marginals, $P(q_O)$, to corresponding inner decoders via the interleaver, $\Pi$, for further decoding rounds. This completes the turbo feedback loop. The outer sheet decoders determine the final marginal outputs, $P(l_O), P(q_O)$, after the iterative feedback and marginal exchanges are completed.}
\label{fig:turbo_decoding_schematic}
\end{figure}

We now develop a decoding method for foliated turbo codes based on the trellis construction methods outlined in section~\ref{sec:fol_conv_trellises} and the SISO decoder in section~\ref{sec:trellis_siso_algorithm}. The trellis construction and SISO decoder allow for the decoding of foliated convolutional codes by iteratively decoding individual sheets followed by a series of marginal exchanges on ancilla qubits. 

Turbo codes consist of an interleaved concatenation of two convolutional codes. The decoding approach is shown schematically in \fig{fig:turbo_decoding_schematic}. Using an \emph{a priori} distribution of qubit error states $\text{Pr}(q_I)$ a soft-input soft-output (SISO) decoder is implemented and marginal values for qubits $P(q_I)$ and logical qubits $P(l_I)$ are calculated for each decoding sheet in the foliated code.  Marginals are then exchanged between the shared ancilla qubits in neighbouring layers and used as prior values for a new round of trellis decoding. This process is applied iteratively.

Within a sheet decoder (i.e.\ red boxes in   \fig{fig:turbo_decoding_schematic}),  the logical marginals, $P(l_I)$, are deinterleaved ($\Pi^{-1}$) and used as priors for the outer decoder, $P(q_O)$. The same process of ancilla marginal exchange is performed and the decoding process is iterated. Finally the qubit marginals from the outer decoder, $P(q_I)$, are interleaved ($\Pi$) and used as logical priors, $P(l_I)$ for the inner code. 

\subsection{Numerical Results for Turbo Codes}
\label{numerical_turbo}

 \tmsA{As noted earlier, $X$ errors on the foliated cluster  commute with parity check measurements.  Thus, for our simulations we assume a phenomenological error model in which uncorrelated $Z$ errors are distributed independently across the  cluster with probability $p$.}   The decoder performance is quantified in terms of both word error rate (WER), which is the probability of one or more logical errors across all $k$ encoded qubits, and the bit error rate (BER) which is the probability of an error in \tmsB{any} of the encoded qubits. \tmsB{These are defined formally in \cref{BERp,WERp}.}

One common approach to numerically evaluating code performance curves is to  sample error patterns, $\vec \varepsilon$, use the decoder to find a recovery operation $\vec e_\text{rec}$, and then test for success or failure of the decoder with respect to the specific error sample.  The decoder is successful if \mbox{$\vec \varepsilon_l\equiv\mathbf{G}.(\vec \varepsilon+\vec e_\text{rec})=\vec 0$}, and unsuccessful if not.   If the decoder fails on any logical qubit, this constitutes a Word Error; the Hamming weight of  $\vec \varepsilon_l$ counts the number of logical Bit Errors.

One approach to generating code performance curves is to fix the error rate per physical qubit, $p$,  then generate $N_\text{trials}$ error configurations at that error rate.  The decoder will fail on some number of those trials, and then the WER  is a function of the error rate, given by 
\begin{equation}
\text{WER}(p) =  \frac{\# \text{Word Failures}}{N_\text{trials}}\big|_{p}.\label{WERp}
\end{equation}  
The error rate is then incremented, $p\rightarrow p'$, and new trials are run for the new value of $p$.

\tmsB{Similarly, we define the BER (at a given error rate, $p$, per physical qubit) to be
\begin{equation}
\text{BER}(p) =  \frac{\# \text{Bit Failures}}{k \,N_\text{trials}}\big|_{p}.\label{BERp}
\end{equation}}

In the numerical results  reported here,  we employ binomial sampling, in which we sample over a fixed number of errors, $j=0,1,2,...$, and compute the failure probability for each $j$  
\begin{equation}
P_{\text{Word}}(j) =  \frac{\# \text{Word Failures}}{N_\text{trials}}\big|_{\text{fixed }j}\label{WERj} 
\end{equation} 
for a suitable range of values of $j$. We then use the binomial formula to relate $\text{WER}(p) $ to $ P_{\text{Word}}(j)$:
\begin{align}
\text{WER}(p) &= \sum\limits_{j=0}^n  P_{\text{Word}} (j) \left(\begin{array}{c}n \\ j \end{array}\right) p^j (1-p)^{n-j}.\label{WERp}
\end{align}
In practice the upper limit of the sum can be truncated to much less than $n$.

Similarly, we define 
\begin{equation}
P_{\text{Bit}}(j) =  \frac{\# \text{Bit Failures}}{k \,N_\text{trials}}\big|_{\text{fixed }j}, 
\end{equation} so that the BER is given by 
\begin{align}
\text{BER}(p) &= \sum\limits_{j=0}^n  P_{\text{Bit}} (j) \left(\begin{array}{c}n \\ j \end{array}\right) p^j (1-p)^{n-j}.
\end{align}

In what follows we present numerical results for code performance  as a function of the code size $k=nr$, and for several different foliation depths, where we vary the number of code sheets, $L$.   We note that the case $L=1$ is a special case: it corresponds to decoding a single clusterised code sheet, in which errors may also occur on the ancilla qubits (i.e.\ the red squares in \fig{codeexamples}).  This is equivalent to decoding the base (i.e.\ unclusterised) code, but including  faulty syndrome measurements.  

As an aside, we validate this binomial sampling method by comparing the  numerical results of \cref{WERj} with numerical results generated by the conventional sampling approach, \cref{WERp}. 
\fig{turbonumericst25} contains a series of WER trials generated using the conventional sampling with for the case of $k= 40$ and $L = 6$ (second panel, LHS), shown as points with error bars, using $N_\text{trials}=10^5$ for the smallest value of $p$. These points are in close agreement with the data generated through binomial sampling (solid lines).

\subsubsection{Numerical Results for T9 Turbo Codes}

\begin{figure}[t]
\begin{center}
\includegraphics[width=\columnwidth]{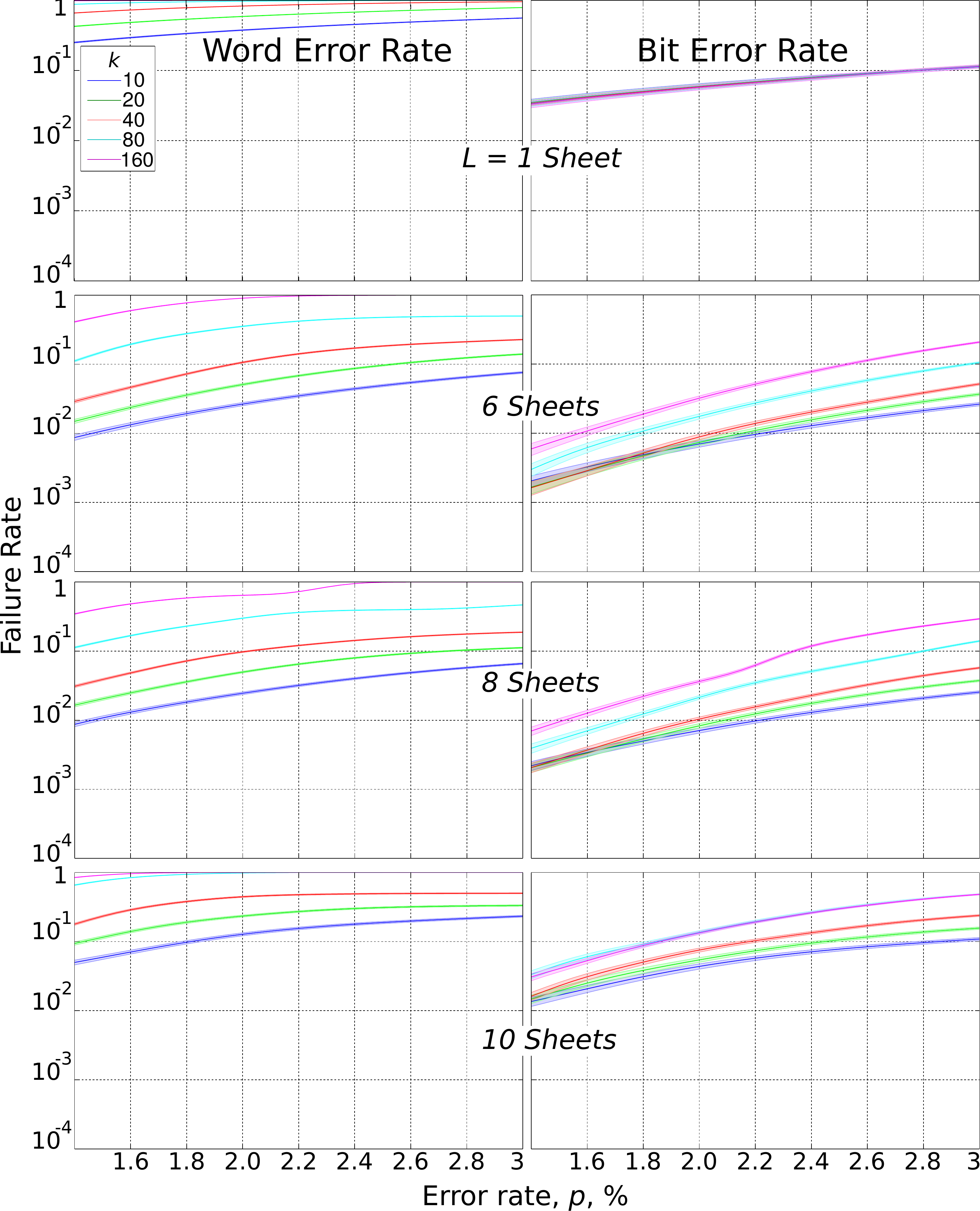}
\caption{Numerical performance results for the foliated T9  \mbox{$[[n,k=n/16,9]]$} turbo code, for different numbers of foliated layers, $L=1,6,8,10$ (rows), \tmsB{as a function of the error rate per qubit, $p$}.  Different colours correspond to different code sizes, $k=n r=10$ (lowest curve),20,40, 80 and $k=160$ (highest curve) logical qubits; shading indicates $\pm1\sigma$. Word Error Rate (left column) counts any error(s) across all $k$ logical qubits.  Bit Error Rate (right column) counts the failure rate per logical qubit.  The T9 code does not exhibit threshold behaviour: its performance degrades as the code size grows.} 
\label{turbonumericst9}
\end{center}
\end{figure}

\fig{turbonumericst9} shows the performance of the  \mbox{$[[n,k=n/16,9]]$} T9 self-dual foliated turbo code.  We see that for a given foliation depth, $L$, the performance \emph{degrades} with the size of the code, $k$.  This indicates that the foliated T9 code is  a poorly performing code.

This is explained by considering the effective distance of the clusterised, and then foliated T9 code. The T9 code is generated using two constituent rate $r=\frac{1}{3}$ C3 codes. By construction these codes have a distance of 3, however when clusterised into a single code sheet  (which is equivalent to accounting for noisy stabiliser measurements that generate faulty syndrome data) the distance is reduced to 2. In this setting the seed generator and stabilizer are
\begin{align}
G &= \left[\begin{array}{C{20mm}C{15mm}C{7mm}|C{4mm}C{4mm}}
$D$ & $D$ & $1$ & $0$ & $0$ 
\end{array}\right],\\
H & = \left[\begin{array}{C{20mm}C{15mm}C{7mm}|C{4mm}C{4mm}}
$1+D+D^2$ & $1+D^2$ & $1$ & $1$ & $1$
\end{array}\right],
\end{align}
 respectively.  (Here, we use the notation that qubits the left of the vertical bar are code qubits in the C3 code, and those to the right of the vertical bar are ancilla qubits associated to code stabiliser measurements; these correspond to quintuplets of qubits  in each frame of the shaded code sheet in \fig{fig:marginal_exchange}b.)
 
For the C3 code, an example of an undetectable weight 2 error pattern  error pattern in a single code sheet is \mbox{$\vec{e} = [0\hspace{2mm}0\hspace{2mm}1\hspace{2mm}|1\hspace{2mm}0 ] $}, which has support on one of the ancilla. Since the distance of the clusterised code sheet is 2, it is reduced to an error \emph{detecting} code within the sheet.  We note that if this specific error pattern were isolated within a larger foliated structure, it \emph{would}
 be detected by parity checks in the adjacent sheets, which have support on the affected ancilla qubit.

 Given the reduced distance of the clusterised code,  we do not necessarily expect this code to perform well in the foliated regime.  This is borne out in the numerical results: for a given foliation depth, $L$; the T9 code has no threshold.

Though this negative result is unsurprising given the foregoing discussion on the clusterised code distance, we show this as an example of a poorly-performing foliated  code.  The simplest way to rectify this issue is to increase the underlying code distance, which we do in the next example.

\subsubsection{Numerical Results for T25 Turbo Codes}

\begin{figure}[t]
\begin{center}
\includegraphics[width=\columnwidth]{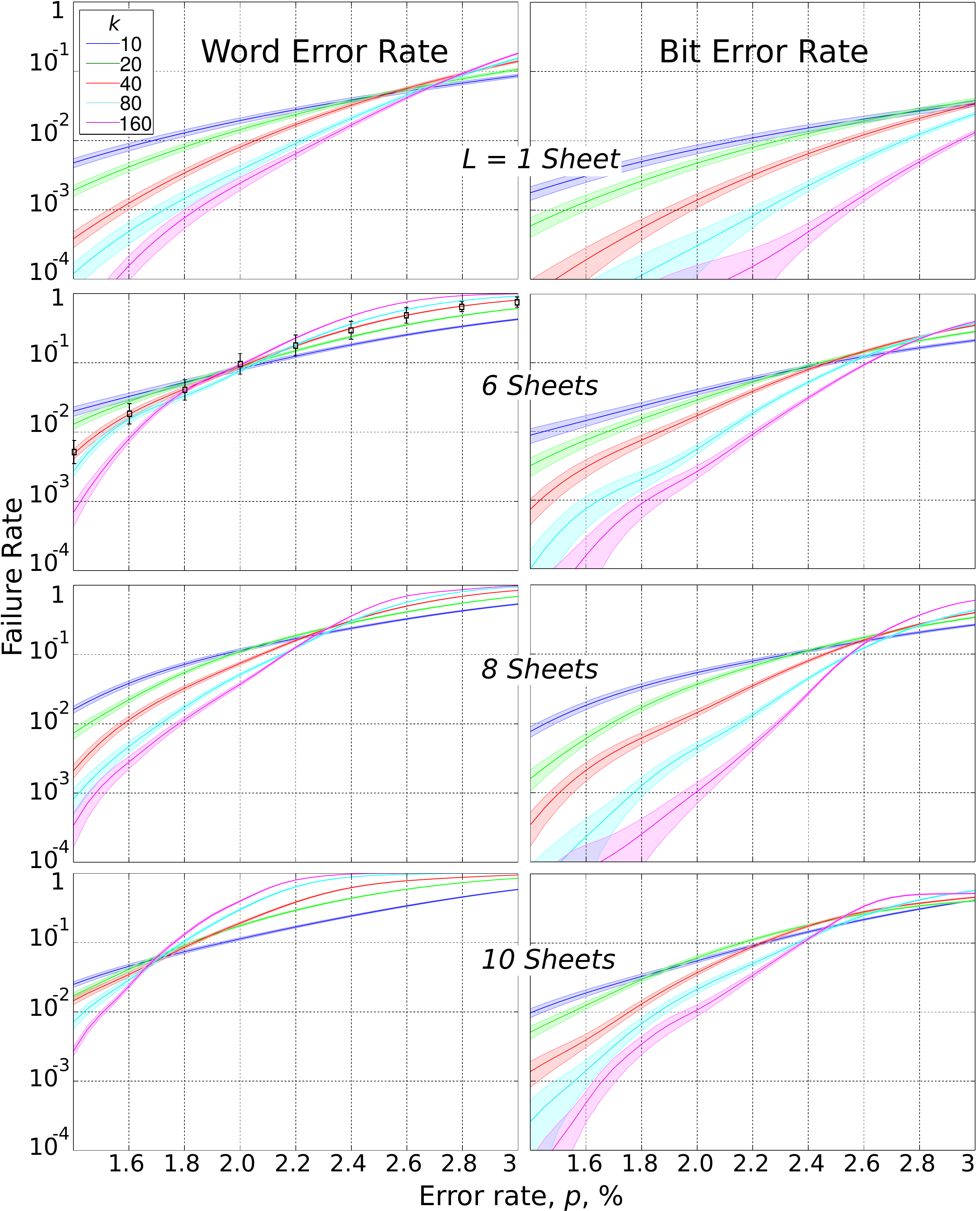}
\caption{Numerical performance results for the foliated T25 \mbox{$[[n,k=n/16,25]]$} turbo code, for different numbers of foliated layers, $L$ (rows), \tmsB{as a function of the error rate per qubit, $p$}.  Different colours correspond to different code sizes with the number of logical qubits indicated as $k=n r=10$ (shallowest curves), 20,40, $80$  and $k=160$ (steepest curves); shading indicates $\pm1\sigma$.  The Word Error Rate (left column) counts any error(s) across all $k$ logical qubits.  Bit Error Rate (right column) counts the failure rate per logical qubit.  The T25 code exhibits threshold-like behaviour, in that for small error rates, the performance of the code improves with code size. In order to verify the binomial sampling process, a series of trials (shown as squares with error bars black for $k=40$, $L =6$) were performed at each value of $p$, and then evaluating \cref{WERp}.} 
\label{turbonumericst25}
\end{center}
\end{figure}

\fig{turbonumericst25} shows the performance of the \mbox{$[[n,k=n/16,25]]$} T25,  self-dual foliated turbo code.

For each $L$, there is a threshold error rate around $p\sim 2\%$, below which the code performance improves with code length (up to  at least 160 encoded logical qubits per code sheet, encoded into 4160 physical qubits per sheet), consistent with (pseudo-)threshold behaviour seen in turbo codes \cite{poulin_2009}.    As $L$ increases, the threshold decreases, more pronouncedly for the WER than the BER.  The range of $k$ and $L$ that we can simulate is limited by computational time, so we cannot explore the asymptotic performance  for   large $L$.  Nevertheless, numerics indicate that foliated turbo codes perform quite well for moderate depth foliations.

 We note  that the foliated construction  transforms a clusterised code into a fault tolerant resource state, but with a reduced threshold.  This is seen in \fig{turbonumericst25}, \tmsB{in which the threshold is seen to reduce with the number of sheets in the foliation}.  A similar effect is also seen in Raussendorf's foliated surface-code construction, in which the fault-tolerant threshold $\lesssim1\%$ is smaller than the $\sim11\%$ threshold \tmsB{(assuming perfect stabiliser measurements)} for the surface code  on which it is based.

One important difference between the surface code and a turbo code family is that the code distance is fixed in the latter, whereas it grows in the former.  As a result, we do not expect this threshold behaviour to survive for large foliation depths: this would be analogous to the degradation in performance of Raussendorf's construction if the transverse code size were held fixed, while the foliation depth were increased.  In the numerical results shown, the code distance is fixed at \mbox{$d=25$}, and so we  expect that for foliation depths $L\gg d$, the threshold will disappear; consequently the code is more correctly described as having a \emph{pseudo-threshold}.  Nevertheless, there may be  applications where this is sufficient for     practical purposes.

\section{Foliated Bicycle Codes}
\label{sec:foliated_bicycle_codes}

\subsection{Construction}
Bicycle code are a class of finite-rate LDPC codes. They are self-dual CSS codes generated by sparse circulant matrices. A circulant matrix is formed by a seed row vector which is rotated by one element in each successive row in the matrix. For a binary sparse cyclical matrix $C = m \times m$ with row weight $w$, a bicycle code can be defined by $H_X = H_Z = [C|C\transpose]$. By construction $H_X$ and $H_Z$ are orthogonal
\begin{align}
H_{X} H_{Z}\transpose &= \big[C\big|C\transpose\big] \big[C\big|C\transpose\big]\transpose = CC\transpose+C\transpose C  = \mathbf{0}.
\end{align}

To create the generator matrix $k$ rows are removed from $H$. This generates a $[[2m,k,d\approx 2w]]$ quantum code. Here the distance is only approximately $2w$ and will depend on the rows removed from $C$ and the construction of $C$ itself.

We can separate the code into two Tanner graph representations, corresponding to $X$ and $Z$ stabilzers. Since bicycle codes are self-dual the Tanner graphs will be identical in both cases.

In the foliated setting stabilisers are parity check operators of the form given in \cref{eq:foliatedparitycheck}. The code can be separated into two Tanner graphs, one which contains qubits within in the primal sub-lattice, and one which contains on qubits in the dual sub-lattice. For example the primal Tanner graph contains the code qubits in odd sheets $2m+1$ and ancilla qubits in even sheets $2m$. Primal parity checks are parity check operators which are centred on the sheets $2m+1$.

\subsection{Decoding}

Bicycle decoding is typically performed using belief propagation on the Tanner graph representation of the code \cite{frey_1998,hagiwara_2007,mackay_1999,mackay_2004}. 
A variable node corresponds to a given physical qubit; and records the likelihood of all possible errors on that qubit. A factor node corresponds to a given stabiliser, which constrains the possible error states of the connected variable (qubit) nodes. 

 A factor graph $G = (V,E)$ is a bipartite graph defined by the set, $V = A\cup I$, of variable nodes $I = \{q_1,q_2,\hdots,q_n\}$, and factor nodes, \mbox{$A = \{a_1,a_2,\hdots,a_{(n-k)/2} \}$}, and edges, $E$, between variable and factor nodes, \mbox{$E=\{(q,a)|a\in A, q\in N(a)\}$}, where $N(a)$ is the set of all variables which appear in constraint $a$.

The belief propagation algorithm calculates marginal distributions for the possible error states of each qubit by using repeated message passing. A \emph{belief}, $b_i(\varepsilon_j)$ represents the probability that qubit $q_i$ has suffered error $\varepsilon_j$; the index $j$ enumerates over possible errors in the error model. This belief is calculated from \emph{messages}, $m_{a\rightarrow q}(\varepsilon_j)\in[0,1]$, in which factor nodes, $a$, report a marginal probability that node $q$ has suffered error $\varepsilon_j$ %
\begin{align}
b_{i}(\varepsilon_j) = \frac{1}{\mathcal{N}_i} \prod\limits_{a\in N(q_i)} m_{a\rightarrow q_i}(\varepsilon_j),
\end{align}
where $\mathcal{N}_i$ is a normalization condition to ensure $\sum_{\varepsilon_j} b_{i}(\varepsilon_j)=1$ at each $q_i$.   

To calculate $b_{i}(\varepsilon_j)$, we also pass messages, $m_{q\rightarrow a}(\varepsilon_j)\in[0,1]$ from qubit nodes to check nodes, reporting the likelihood that $q$ is subject to error $\varepsilon_j$.  The values of messages in both directions are determined by iterating over the 
 following consistency conditions
\begin{align}
m_{a\rightarrow q}(\varepsilon_j) &= \hspace{-4mm}\sum\limits_{\vec{e}_p: p\in N(a)\backslash q}\hspace{-4mm}f_a(\vec{e}_p|\varepsilon_j\text{ on }q)\hspace{-3mm}\prod\limits_{r\in N(a)\backslash q}\hspace{-4mm}m_{r\rightarrow a}({e}_{p,r}),\label{eq:BP1}\\
m_{q \rightarrow a}(\varepsilon_j) &= \prod\limits_{b\in N(q)\backslash  a}m_{b\rightarrow q}(\varepsilon_j).
\label{eq:BP}
\end{align}
Here we sum over all possible configurations of errors, $\vec{e}_p$, over the neighbours of $a$ (excluding the target qubit $q$), and 
 $f_a(\vec{e}_p|\varepsilon_j\text{ on }q)\in[0,1]$ are constraint functions that return the \emph{a priori} likelihood of the error configuration $\vec{e}_p$ given that qubit $q$ is subject to error $\varepsilon_j$, and ${e_{p,r}}$ is the restriction of the error configuration $\vec{e}_p$ to qubit $r$. The function $f$ serves two purposes: it vanishes on error configurations that are inconsistent with syndrome data, and otherwise returns the likelihood of a valid error configuration.  For our numerical simulations, we will assume independently distributed errors, which implicitly defines $f$.  We note in passing that $f$ may be tailored to correlated error models if necessary.
 
To begin the iterative message passing, we initialise the messages on the RHS of \cref{eq:BP1} using the \emph{a priori} error model 
\begin{align}
m_{q \rightarrow a}(\varepsilon_j) &=\text{Pr}(\varepsilon_j).
\end{align}

Belief propagation is exact on tree graphs, allowing for the factorisation of complete probability distribution into marginals over elements, and Equations (\ref{eq:BP1}) and (\ref{eq:BP}) naturally terminate at the leaves of the tree \cite{mackay_1997}.  For loopy graphs, the iterative message passing does not terminate in a fixed number of steps; rather we test for convergence of the messages to a fixed point.  In some cases, particularly at higher error rates, the message passing may not converge; in this case we simply register a decoding failure on the subset of logical qubits that are affected.  Further, the presence of many short cycles may lead to poor performance of the decoder. In the foliated bicycle code, there are numerous short, inter-sheet, graph cycles, however in the numerical results  we present in the next subsection, we see empirically that the decoder is effective nonetheless. 

Finally, we note that the message passing algorithm described here can be applied directly to the full foliated code.  Here, the marginal exchange between sheets happens concurrently with intra-sheet message passing, i.e.\ \cref{eq:BP} performs both inter-sheet marginal exchange and intra-sheet message passing.

\subsection{Numerical Results for Bicycle Codes}
\label{sec:numerical_results_bic}

 \begin{figure}[t]
\begin{center}

\includegraphics[width=\columnwidth]{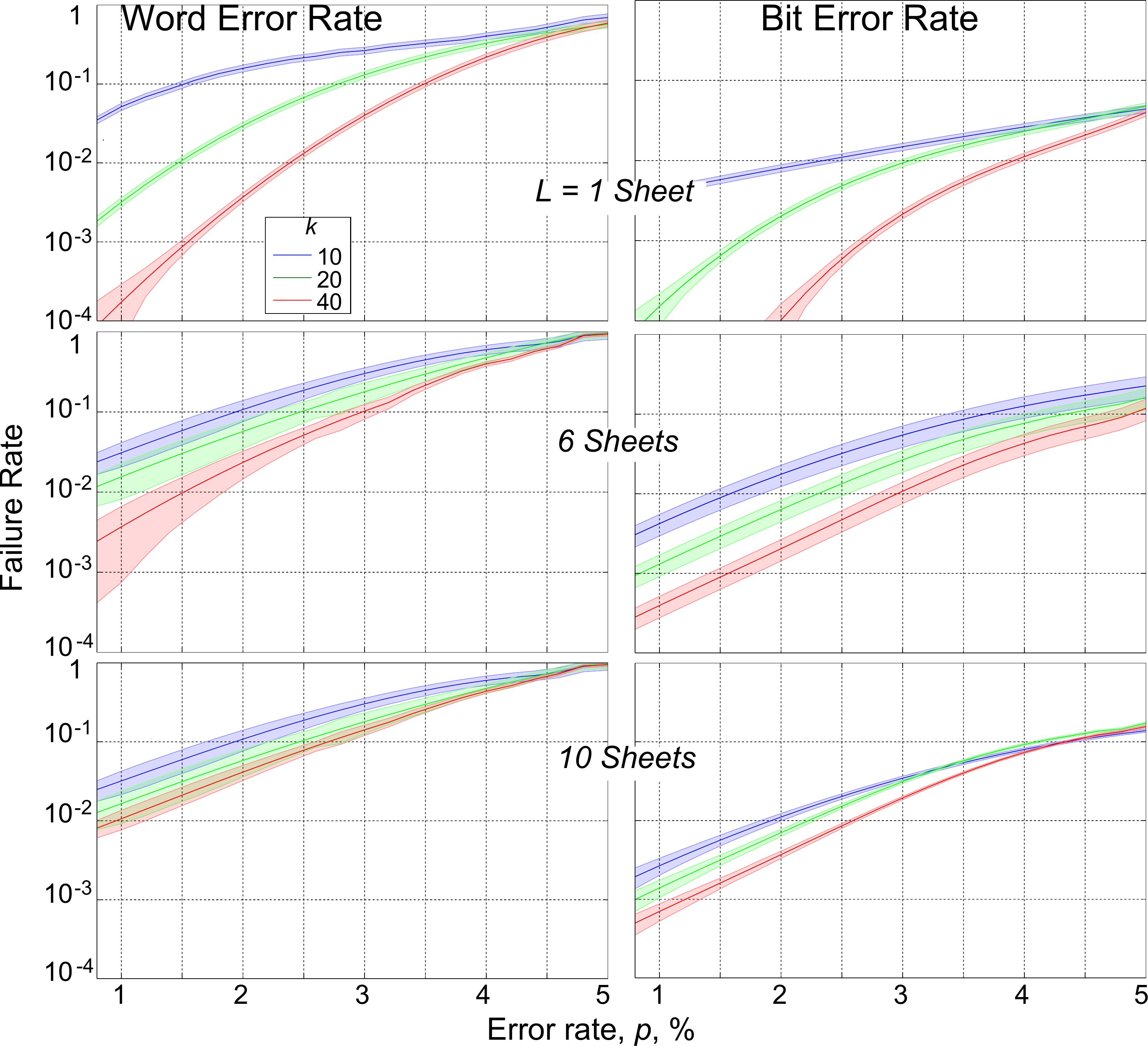}
\caption{Numerical performance results for a foliated \mbox{$[[n,k=n/16,d\approx26]]$} bicycle code, for different numbers of foliated layers, $L=1,6,10$ (rows), \tmsB{as a function of the error rate per qubit, $p$}.  Different colours correspond to different code sizes, $k=n r=10$ (shallowest curves) $k=20$, and $k=40$ (steepest curves); shading indicates $\pm1\sigma$.  Word Error Rate (left column) counts any error(s) across all $k$ logical qubits.  Bit Error Rate (right column) counts the failure rate per logical qubit.} 
\label{bicnumerics}
\end{center}
\end{figure}

We analyse the performance of the codes as a function of the code size $k=nr$, and the number of foliated layers, $L$. \fig{bicnumerics} shows the performance of a $d\leq 26$,\footnote{this distance bound is established by a heuristic minimisation of the length of the logical operators.}  $r={1}/{16}$,  bicycle code, based on Monte Carlo  simulations of errors.  We use the same binomial sampling schedule as described in Section \ref{numerical_turbo}.  

For each $L$, there is a threshold error rate around $p\sim 4.5\%$, below which the code performance improves with code length (up to  at least 40 encoded logical qubits per code sheet). As with the Turbo codes, as $L$ increases, the performance increase gained from larger codes is diminished.

This result shows that LDPC bicycle codes are potentially promising  codes for foliating.

\section{Clusterised Code Architecture}
\label{sec:turbo_code_arch}

In this section we analyse the gate based implementation of cluster state resources for clusterised convolutional and turbo codes. The faulty implementation of cluster bonds (\textsc{c-phase} gates) causes correlated errors to arise during the construction of the cluster state resource. It is important to model these types of errors and ensure that the decoding process is fault-tolerant.

\subsection{Schedule for Cluster-state construction}

Cluster state construction requires the preparation of resource $\ket{+}$ qubits and the implementation of \textsc{c-phase}  gates, $\Lambda(a,b)$, between pairs of qubits. The gates must be implemented over a series of time steps so that during any given time step, $T_j$, any qubit is addressed by at most one phase gate. To generate clusterised codes phase, gates are implemented between ancilla qubits and code qubits according to the Tanner graph of the $\mathcal{S}_Z$ stabilisers.  The number of ancilla qubits is $|\mathcal{S}_Z|$. The minimum number of time steps required to implement pairwise \textsc{c-phase} gates between each ancilla and the code qubits is proportional to the weight of the largest stabiliser.

The bonds in a clusterized convolutional code can be characterised by a series of qubit pair operations $\Lambda_{T_m}(a_{i,j},c_{i',j'})$, where $a_{i,j}$ refers to the $i$th ancilla qubit in frame $j$, $c_{i',j'}$ refers to the $i'$th code qubit in frame $j'$ and $T_m$ is a `time' index in the cluster construction schedule.

\begin{figure}[t]
\includegraphics[width=\columnwidth]{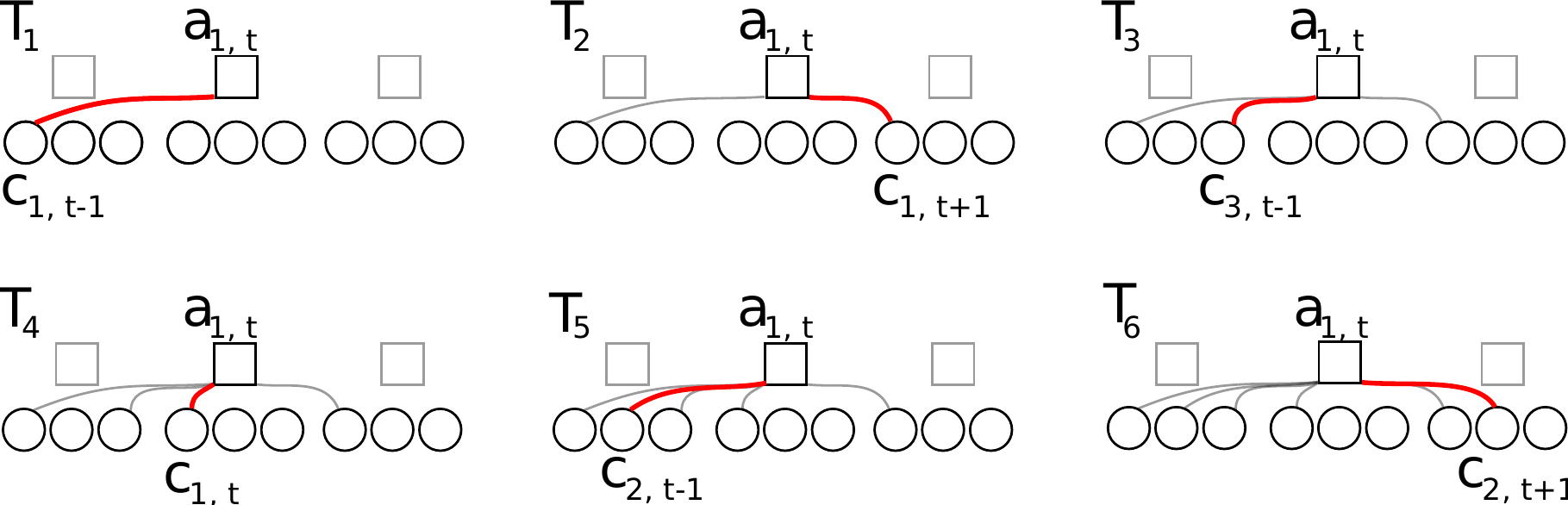}
\caption{A suitable time ordering for implementing \textsc{c-phase} gates between ancilla qubits and code qubits for the C3 convolutional code defined in \cref{eq:example_conv}, using the transpose interleaver defined in \cref{eq:pvec_pi}. Translations of these gates generate the full clusterised code.  Thick, red lines indicate  \textsc{c-phase}  to implement at the corresponding time $T_j$; fine grey lines indicate previously implemented gates. The distance between frames $t$ and $t'$, and frames $t'$ and $t''$ are $\frac{\tau}{3}$.}
\label{fig:c3_time_ordering}
\end{figure}

As an example consider the C3 code defined in equations~\ref{eq:example_conv}. The parity check operators are weight 6, and the corresponding ancilla qubit are represented by square vertices in the figure.  As a result, this clusterised code may be implemented in 6 time steps, $T_1, ..., T_{6}$.  A possible schedule for implementing \textsc{c-phase} gates is shown in \fig{fig:c3_time_ordering}. Thick, red lines indicate  \textsc{c-phase}  to implement at the corresponding time $T_j$; fine grey lines indicate previously implemented gates.  Implied, but not shown are simultaneous translations of these gates across all frames, indexed by $..., t-1,t,t+1,...$.

For the case of Turbo codes, the number of cluster bonds between an ancilla qubit for an outer parity check depends on the weight of the outer convolutional parity check, the weight of the inner generator, and the choice of interleaver. The outer parity check is formed by encoding using the inner code generator. In the case of a random interleaver and a large code each bit within the convolutional parity check is distant from each other bit. As a result the total weight for the outer parity check will be $\text{wt}(H_{\text{outer}}) \times \text{wt}(G_\text{{inner}})$. 

{
\addtocounter{footnote}{-2}
\renewcommand*{\thefootnote}{\fnsymbol{footnote}}

As we discussed in section~\ref{turboconstruction}, the choice of interleaver has an effect on the weight of the code stabilisers.  In the T9 and T25 clusterised turbo codes, a completely random interleaver will generate stabilisers with weights up 18\footnote{\hspace{-1mm}\mbox{$\text{wt}(G_{\text{inner}}) \times \text{wt}(H_{\text{outer}}) = 3\times 6.$}} and 98\footnote{\hspace{-1mm}\mbox{$\text{wt}(G_{\text{inner}}) \times \text{wt}(H_{\text{outer}}) = 7\times 14.$}} respectively.} In what follows,  we describe more structured interleavers that reduce these weights to 10 and 26 respectively\footnote{\tmsB{Note that we have not done threshold simulations for thes einterleaver; the numerical results presented in \fig{turbonumericst25} may depend on the choice of interleaver.}}.

\begin{figure}[t]
\includegraphics{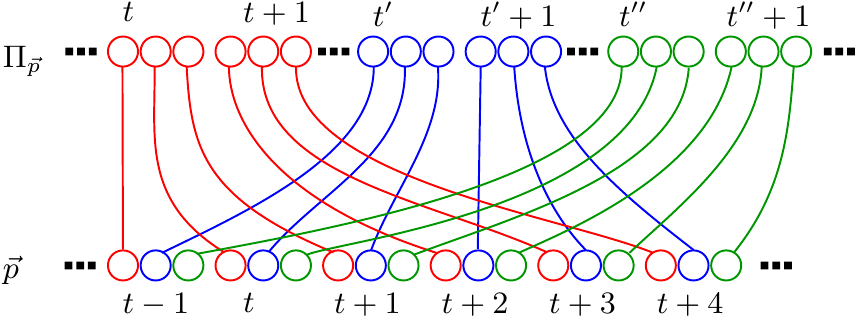}
\caption{The transpose interleaver for a rate $\frac{1}{3}$ code. The interleaver takes a sequence of frames containing bits, $\vec p$, and returns a sequence, $\Pi_{\vec p}$, containing the first bits in every frame, followed by the second bits in every frame and then the third bits in every frame. If $\vec{p}$ contains $\tau$ frames then $t'=t+\tau /3 $ and $t''=t+2\tau /3 $.}
\label{fig:frame_position_interleaver}
\end{figure}

An interleaver that achieves these lower weight stabilisers is one which permutes the order of bits $\vec{p}$ such that in a block of $f$ frames, the first bit within in each frame is mapped to a single contiguous block $\vec b_t$ by the permutation; the second bit within each frame is mapped to another, well spaced block, $b_{t'}$, and so on.  
 This is illustrated in \fig{fig:frame_position_interleaver}. The  input sequence over $\tau$ frames is
\begin{align}
\vec{p} 
&= ((p_1^1,p_1^2,...,p_1^n),(p_2^1,...,p_2^n),...,(p_\tau^1,...,p_\tau^n))\nonumber\\
&\equiv (\vec{p}_{1},\vec{p}_{2},...,\vec{p}_{\tau})
\label{eq:pvec}
\end{align}
where $\vec{p}_{t} = (p^1_t,p^2_t,\hdots, p^n_t)$ is the input vector over frame $t$, with $n$ physical (qu)bits. The interleaver, $\Pi$,  applies a permutation on $\vec{p}$ such that 
\begin{align}
\Pi_{\vec p} &= ((p^1_1,p^1_2,\hdots,p^1_\tau),(p^2_1,\hdots,p^2_\tau),...,(p^n_1,\hdots p^n_\tau))\nonumber\\
&=(\vec b_1,\vec b_2,...,\vec b_n)\label{eq:pvec_pi}
\end{align}
We  call this interleaver a \emph{transpose} interleaver\footnote{if we write $\vec{p}$ as a $\tau\times n$ matrix where the $\vec{p}_t$ are row vectors, then $\Pi_{\vec p}=\vec p\transpose$ is the $n\times \tau$ matrix transpose.}. This interleaver does not disperse bits as widely throughout the bitstream as a completely random interleaver, however it does generate inner parity check stabilizers which have significantly lower rate than $\text{wt}(H_{\text{outer}})\times \text{wt}(G_{\text{inner}})$.

Using a transpose interleaver, the cluster construction schedule is shown in \fig{fig:time_ordering} for the T9 code.  In this example, there are two classes of parity check operators for each frame: square vertices correspond to weight 6 stabilisers, and the diamond vertices correspond to weight 10 stabilisers.  As a result, this clusterised code may be implemented in 10 time steps, $T_1, ..., T_{10}$.

Also using a transpose interleaver, the T25 code can be implemented in 26 time steps, and has maximum weight 26 stabilisers.  We present the cluster-state construction schedule and interleaver details in appendix \ref{T25schedule}.

To generate the cluster resource for a foliated code each sheet can be generated independently using a suitable schedule for the corresponding primal and dual clusterised codes. An additional two time-steps are then required to connect the code qubits of neighbouring sheets to build the fully foliated network.

\subsection{Error propagation during cluster-state construction schedule}
\label{scheduleerror}

During the construction of cluster states errors may accumulate on individual qubits, or be caused by faulty gate implementation between qubits. We simplify the analysis of these errors by restricting our analysis to $X$ and $Z$ Pauli errors. A $Z$ error commutes with \textsc{c-phase}  gates, however an $X$ error does not, and will propagate a $Z$ error to the neighbouring qubit.

Consider the case where an $X$ error occurs on an ancilla qubit, $a_k$, at some time, $T_\varepsilon$ in the construction schedule.  Subsequent \textsc{c-phase} gates will generate $Z$ errors on all code qubits $c_i \in N(a_k)$ subject to gates $\Lambda_{T_k}(c_i,a_k)$ where $T_k>T_\varepsilon$. Note that  $\bigotimes_{c_i \in N(a_k)} Z_{c_i}$ is a  stabiliser, so the this error pattern is equivalent $Z$ errors on  qubits $c_i$ subject to gates $\Lambda_{T_{k'}}(c_i,a
_k)$ where $T_{k'}\leq T_\varepsilon$. As a result, the maximum number of $Z$ errors arising during the cluster construction is equal to half the weight of the stabiliser. For this reason codes with low weight parity checks are  desirable.

\begin{figure}[t]
\includegraphics[width=\columnwidth]{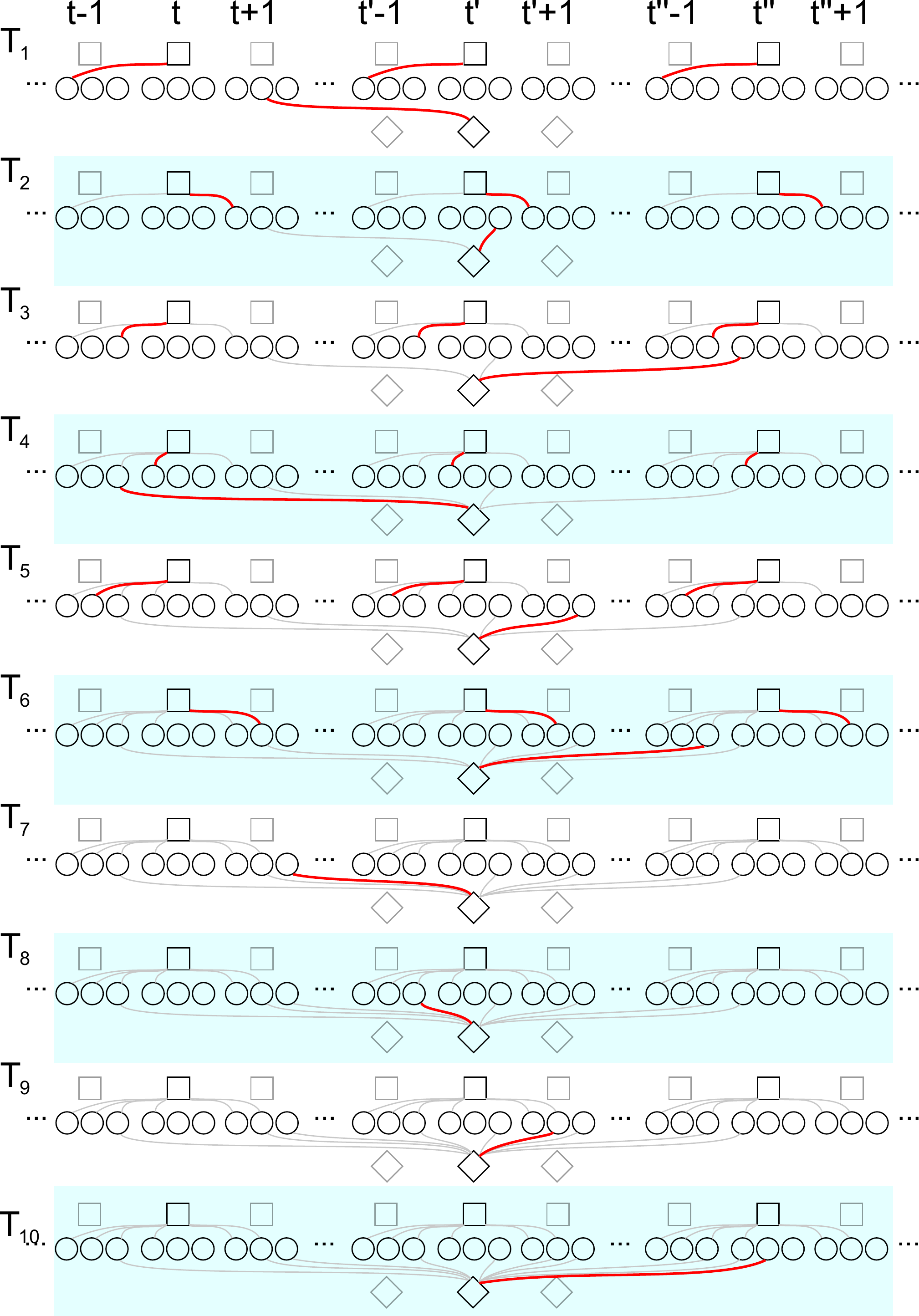}
\caption{A suitable time ordering for implementing \textsc{c-phase}  gates between ancilla qubits and code qubits in a clusterised T9 turbo code, which is a concatenation of two C3 convolutional codes. The gates for 3 inner seed stabilisers (square ancillas) and one outer seed stabiliser (diamond ancilla) are shown. All other stabilisers are translations of these. Each qubit is acted on by at most one gate at each time step. Thick, red lines indicate  \textsc{c-phase}  to implement at the corresponding time $T_j$, fine grey lines indicate previously implemented gates.}
\label{fig:time_ordering}
\end{figure}

The choice of time ordering for gates affects the types of correlated error patterns that arise during construction. Depending on the choice of code, some time orderings may be more favourable than others, producing error patterns which are more likely to be corrected.  One criterion that should be met is that any single physical error during the cluster construction should lead to a correctable (i.e.\ decodable) correlated error pattern after the $\Lambda$ gates have been made. If the number of correlated errors is less than half the code distance, $w_\text{max} < d/2$, then this condition is always met. On the other hand, if the number of errors which are propagated is larger than $d/2$, then we must verify explicitly that the resulting error pattern is correctable.

As an example, the  schedule for constructing the $d=9$ T9 code, which is shown in \fig{fig:time_ordering} using the transpose interleaver.  This code has stabilisers 
of weight 10 associated to the ancillae indicated by diamonds.  As a result, an $X$ error midway through the cluster construction could result in a pattern of up to $w_\text{max}=5$ correlated $Z$ errors on code qubits adjacent to the diamond ancillae.  In this case, even though $w_\text{max}>d/2$, we have checked that the decoder correctly corrects all such errors arising in the schedule in \fig{fig:time_ordering}. 

Similarly, for the $d=25$ T25 code, a maximum stabilizer weight of 26 can be achieved using the transpose interleaver. A physical error during cluster construction could cause correlated error of weight $w_\text{max}=13$.  Again, In this case, even though $w_\text{max}>d/2$, we have checked that the schedule  for cluster construction (listed in Appendix \ref{T25schedule}) produces an error pattern which is correctly decoded.

\section{Conclusion}


In conclusion, we have shown how to clusterize arbitrary CSS codes.  We have shown how to foliate  clusterised codes, generalising Raussendorf's 3D foliation of the surface code.  We have described a generic approach to decoding errors that arise on the foliated cluster using an underlying soft decoder for the  CSS code as a subroutine in a BP decoder, and applied it  to  error correction \tmsA{by means} of a  foliated turbo code. We have also shown how decoding can be performed in the case of foliated bicycle codes.  These construction may have  applications where codes with finite rate are useful, such as long-range quantum repeater networks.

We have exemplified the foliated construction with several code families, namely the T9 and T25 codes, and a the LDPC Bicycle code.  \tmsB{We believe this is the first example of a finite rate generalisation of a sparse cluster-state code with pseudo-threshold behaviour (i.e.\ up to moderately large code sizes)}.  The T25 and the Bicycle code both exhibit threshold-like behaviour, even with moderate levels of foliation.

An important direction for future work is the analysis of error models that take into account the cluster state construction schedule in section \ref{scheduleerror}, the correlated error patterns that arise during construction.  Another direction that will be important for repeater application is the tolerance of the foliated  construction and decoding process to erasure errors.  This is likely to depend on the percolation threshold in the corresponding tanner graph, as discussed in \cite{barrett_2010}.  Finally, developing code deformation protocols for performing gates within  the general foliated architecture is an important direction for future research.

\acknowledgments{
This work was  funded by ARC Future Fellowship FT140100952 and the ARC Centre of Excellence for Engineered Quantum Systems CE110001013, \tmsB{NSERC and the Canadian Institute for Advanced Research (CIFAR)}. 
 We  thank Sean Barrett, Andrew Doherty, Guillaume Duclos-Cianci, Naomi Nickerson, Terry Rudolph, Clemens Mueller and Stephen Bartlett for helpful conversations.
}

\appendix

\section{Example of Inverse  Syndrome Formers}
\label{ISFexample}

\tmsB{We provide an example of the construction of an Inverse Syndrome Former (ISF) and the corresponding pure errors, using the example of the 7-qubit Steane code.
asad}

\tmsB{The Steane code is self dual, so that the generator, $\mathbf{G}$, parity check, $\mathbf{H}$ and $\mathbf{ISF}$ matrices are the same for the $X$- and $Z$-like operators, so we will drop Pauli labels.  The parity check matrix $\mathbf{H}$ for the Steane code is given by the binary support vectors corresponding to the stabilisers defined in $\mathcal{S}_Z^{\text{Steane}} $ in \cref{stabilisers}, and $\mathbf{G}\transpose=\{1,1,1,1,1,1,1\}$.  The $\mathbf{ISF}$ is chosen to satisfy \cref{eq:commutation_conditions}. We group $\mathbf{G}$, $\mathbf{H}$    and one possible choice of $\mathbf{ISF}$  as sub-matrices in a composite, square matrix
{\renewcommand\arraystretch{1.05}
\begin{equation}
\left[\begin{array}{c}
\mathbf{G}\transpose\\
\hline
\mathbf{H}
\vphantom{\left[\begin{array}{c}1\\ 1\\ 1 \end{array} \right]}\\
\hline
\textbf{ISF} 
\vphantom{\left[\begin{array}{c}1\\ 1\\ 1 \end{array} \right]}\\
\end{array} \right]
=\left[\begin{array}{ccccccc}
1&1&1&1&1 &1&1
\vphantom{\mathbf{G}\transpose}
\\  
\hline
1&1&0&0&0 &1&1\\  
0&1&1&1&0 &0&1\\  
0&0&0&1&1 &1&1\\  
\hline
 0 & 1 & 1 & 0 & 0 & 0 & 0 \\
 1 & 1 & 0 & 0 & 0 & 0 & 0 \\
 0 & 0 & 1 & 1 & 0 & 0 & 0    
 \end{array} \right].
\end{equation}\renewcommand\arraystretch{1}}
It is straightforward to check that,
\begin{equation}
\left[\begin{array}{c}\mathbf{G}\transpose\\\mathbf{H}\\ \textbf{ISF} \end{array} \right].
\left[\begin{array}{ccc}\mathbf{G} & \textbf{ISF}\transpose  &\mathbf{H}\transpose  \end{array} \right]=\mathbb{I}_{7\times 7},
\end{equation}
consistent with \cref{eq:commutation_conditions}, i.e.\ $\textbf{ISF}\transpose$ is a pseudo-inverse to $\mathbf{H}$.}

\tmsB{A given error chain, $\vec\varepsilon$, yields a syndrome \mbox{$\vec S=\mathbf{H}.\vec\varepsilon\in\mathbb{Z}_2^3$}, from which we can compute a pure error  \mbox{$\vec{e}^{\,0}=\textbf{ISF}\transpose.\vec S$}.  Since  $\mathbf{H}.\textbf{ISF}\transpose=\mathbb{I}$ the pure error  satisfies \mbox{$\mathbf{H}.\vec{e}^{\,0}=\mathbf{H}.\textbf{ISF}\transpose.\vec S=\vec S$}, i.e.\ it has the same syndrome as the actual error.  Equivalently, $\vec\varepsilon+\vec{e}^{\,0}$ is a logical operator on the code space.}

\section{Transfer Function Notation}
\label{ap:trans_func_notation}

Transfer functions are a convenient method of expressing codes families which have a regular structure but an arbitrary length. Convolutional codes are a family of codes which are often represented by transfer functions. In this appendix we show the relationship between the full matrix expressions for the generators, $\mathbf{G}$, and the seed generators, $G$.

A $k \times 1$ vector of logical bits is expressed by vector $L\transpose = \left[ l_1, l_2,\hdots, l_k\right]$. The generator matrix $\mathbf{G}$ can be represented using a finite dimensional \emph{seed} generator with delay operations. We introduce $\mat{D}$ and $\matconj{D}$ which are $f\times fn$ and $fn \times f$ matrices defined by $\mat{D} = \left[\begin{array}{c|c|c|c} 
D^0\mathbb{I}_f & D^1\mathbb{I}_f & \hdots & D^{k-1}\mathbb{I}_f 
\end{array}\right]$ and $\matconj{D}\transpose = \left[\begin{array}{c|c|c|c} 
\widetilde{D}^0\mathbb{I}_f & \widetilde{D}^1\mathbb{I}_f & \hdots & \widetilde{D}^{k-1}\mathbb{I}_f 
\end{array}\right]$ for a rate $\frac{1}{f}$ code. The operators $D$ and $\widetilde{D}$ satisfy $\widetilde{D}^i \times D^j = \delta_{ij}$. Here $D$ is the usual delay operator and $\widetilde{D}^i$ is a mnemonic for the inverse of $D^i$. Then we have
\begin{align}
\matconj{D} \times \mat{D} &= \left[\begin{array}{c|c|c}
\widetilde{D}^0D^0\mathbb{I}_f  & \hdots & \widetilde{D}^0D^{k-1}\mathbb{I}_f \\
\widetilde{D}^1D^0\mathbb{I}_f  & \hdots & \widetilde{D}^1D^{k-1}\mathbb{I}_f \\
\vdots &   & \vdots\\
\widetilde{D}^{k-1}D^0\mathbb{I}_f  & \hdots & \widetilde{D}^{k-1}D^{k-1} \mathbb{I}_f
\end{array}\right] \mathbb{I}_{fk}.
\end{align}
We write $\mathbf{G}L = \matconj{D}\,\mat{D}\mathbf{G}L$ from which we can derive finite size seed generator.  

As an illustration consider a rate $\frac{1}{3}$ code with generator matrix
\begin{align}
\mathbf{G}\transpose &= \left[
\begin{array}{cccccc}
111    & 100          & 110       &             &        &          \\
       & 111          & 100       & 110         &        & \hdots          \\
       &              & 111       & 100         & 110    &   	
\end{array} \right].
\label{eq:conv_block_generator_example}
\end{align}
We can express the encoding as
\begin{align}
&\mathbf{G}L=\matconj{D}\left[\begin{array}{c|c|c}\mathbb{I}_f & \mathbb{I}_f D & \hdots \end{array}\right]\times \mathbf{G} \times
\left[\begin{array}{c}
l_1\\
l_2\\
\vdots\\
l_k
\end{array}\right],
\\
 &= \matconj{D}\left[{\medmuskip=-2mu
\thinmuskip=-2mu
\thickmuskip=-2mu
\nulldelimiterspace=-1pt\begin{array}{cccc}
1+D+D^2 & D+D^2+D^2 & D^2+D^3+D^4  &\\
1+D^2   & D+D^3     & D^2+D^4      &\hdots \\
1       & D         & D^2         &
 \end{array}} \right]\hspace{-1mm}\left[\begin{array}{c}
l_1\\
l_2\\
\vdots\\
l_k
\end{array}\right]\nonumber\\
&= \matconj{D}\left[\begin{array}{c}
(1+D+D^2)(l_1+l_2D+\hdots + l_kD^{k-1}) \\
(1+D^2)(l_1+l_2D+\hdots + l_kD^{k-1})   \\
(1) (l_1+l_2D+\hdots + l_kD^{k-1})
\end{array} \right],\nonumber\\
&= \matconj{D} \left[\begin{array}{c}
1+D+D^2\\
1+D^2\\
1
\end{array} \right] \sum_{i=0}^{k-1} D^i l_{i+1},\nonumber\\
&= \matconj{D}G \sum_{i=0}^{k-1} D^i l_{i+1},
\end{align}
where $\matconj{D}$ is a $fn\times f$ matrix, with $f=3$. Here we have $D^0 \equiv 1$. In the last line we define the seed generator, $G$, which is a $3 \times 1$ matrix defined in terms of delay operators.

To output the physical qubits from this seed generator a string of logical input bits $L$ is multiplied by $\Delta(D) = [1 \,D \,D^2 \hdots]$. In our working example we have
\begin{align}
\sum_{i=0}^{k-1} D^i l_{i+1} &\equiv \Delta(D) \times L.
\end{align}
The code qubits can be determined by multiplying this expression by $G$. For the more general case of a rate $\frac{b}{f}$ code takes $b$ logical inputs and produces $f$ physical outputs at each frame. For an encoding operation we have
\begin{align}
c &= GL, 		\nonumber\\
  &= \matconj{D}\hspace{0.5mm}\mat{D} G L, \nonumber\\
	&= \matconj{D}G\Delta L,
\end{align}
where $[c] = fn\times 1$, $[\matconj{D}] = fn \times f$, $[\mat{D}] = f \times fn$, $[G] = f \times b$ and $[\Delta] = b \times bn$.

\section{Transfer Function Manipulation}

Now that we have established the implementation of transfer functions as a description of convolutional codes consider the problem of determining the $\textrm{ISF}$. The standard approach takes a pseudo inverse of the seed generator matrix. For a rate $\frac{b}{f}$ code the generator matrix has size $nb \times nf$, where $n$ is the number of frames. We have the property
\begin{align}
\mathbf{G}^{-1} \times \mathbf{G} = \mathbb{I}_{nb\times nb}.
\end{align}
To express this in terms of transfer function notation we have 
\begin{align}
\mathbb{I}_{nb\times nb} &= \matconj{D}\mathbf{G}^{-1}\mathbf{G}\mat{D},\nonumber\\
             &= \matconj{D}\mathbf{G}^{-1} G(D) \left[{\medmuskip=-2mu
\thinmuskip=-2mu
\thickmuskip=-2mu
\nulldelimiterspace=-1pt \begin{array}{cccccccc}
             1 & D & \hdots & D^{k-1} & 0 & 0 & \hdots & 0             \\
             0 & 0 & \hdots & 0       & 1 & D & \hdots & D^{k-1}       \\
               &   &        &         & \vdots  &   &        &    
             \end{array}}\right],\nonumber
						\end{align}
						\begin{align}            
						&=\Delta\transpose(\widetilde{D}) G^{-1}(\widetilde{D})G(D)\Delta(D).\label{eq:delta_D_generator_identity}
\end{align}
We can use the identity 
\begin{align}
\widetilde{D}^jD^{i} = \delta^{ij} = \widetilde{D}^0D^{i-j},
\label{eq:multiplication_law}
\end{align}
to express functions of $\widetilde{D}$ in terms of $D$. Note that the order of operations must be performed so that all $D$ follow $\widetilde{D}$ terms. From \cref{eq:delta_D_generator_identity} we have
\begin{align}
\mathbb{I}_{nb} &= \widetilde{D}^0\Delta\transpose(D^{-1})G^{-1}(D^{-1})G(D)\Delta(D),\nonumber\\
                &= \widetilde{D}^0  \Delta{D}    \Delta\transpose(D^{-1})G^{-1}(D^{-1})G(D)\Delta(D) \Delta\transpose(\widetilde{D}),\nonumber\\
								&= \widetilde{D}^0  G^{-1}(D^{-1})G(D),\nonumber\\
								&= G^{-1}(\widetilde{D}) G(D).
\end{align}

The pseudo inverse of $G(D)$ is $G(D^{-1})$. To calculate this we make a small alteration to $G(\widetilde{D})$ by substituting the terms $D^i$ for $\widetilde{D}^{-1}$.

We now work through an example to demonstrate this approach. Consider the case of a rate $\frac{2}{3}$ convolutional code where we wish to find its parity checks matrix. One of the generators is taken from our working example and the second generator input is $\left[D, D, 1 \right]$. We have
\begin{align}
[G\transpose|I] = \left[
\begin{array}{ccc|ccc}
D & 1 & 1 & 1 & 0 & 0\\
    1+D+D^2   &  1+D^2    & 1 & 0 & 1 & 0\\
        &       &     & 0 & 0 & 1
\end{array}
\right],
\end{align}
has a pseudo inverse of 
\begin{align}
[I|G^{-1}] = \left[
\begin{array}{ccc|ccc}
1       & 0     & 0 & 1+D       & 1+D  & 1+D+D^2  \\
0       & 1     & 0 & D         & D    & 1+D^2    \\
        &       &   & 1+D       & D    & D^2
\end{array}
\right].
\label{eq:pseudo_conv_inverse1}
\end{align}
To satisfy the condition $G^{-1}G = \mathbb{I}$ we note that the product of the first column of $G^{-1}$ and the first row of $G$ must be 1. The same is true for the second column and second row. The product of the third column of $A$ and the generators must be zero. Since $HG = 0$, and the number of parity checks is $n-k=1$, this means the third column must be equivalent to $H\transpose$. Expressed in transfer function form this gives us
\begin{align}
H(D^{-1}) = \left[
\begin{array}{C{2.2cm}C{1.5cm}C{1cm}}
$1+D+D^2$&$1+D^2$&$D^2$
\end{array}
\right].
\end{align}
It follows that
\begin{align}
H(D) &= \left[\begin{array}{C{2.2cm}C{1.5cm}C{1cm}}$1+D^{-1}+D^{-2}$ & $1+D^{-2}$ & $D^{-2}$  \end{array}\right],\\
&=\left[
\begin{array}{C{2.2cm}C{1.5cm}C{1cm}}
$1+D+D^2$   & $1+D^2$ & $1$
\end{array}
\right]\times D^{-2}.\nonumber
\end{align}
We recognise that the parity check is equivalent to the first seed generator shifted by 2 frames. If we compensate for the shift then the two are equivalent ($H_X = H_Z$) since this convolutional code is is self-dual. Performing an inverse on \cref{eq:pseudo_conv_inverse1} we should reclaim the original seed generators as well as the inverse syndrome former.
\begin{align}
\hspace{-3mm}\left[{\medmuskip=-2mu
\thinmuskip=-2mu
\thickmuskip=-2mu \begin{array}{c|c}G_0\transpose(D^{-1})&I\end{array}}\right] &= \left[
{\medmuskip=-2mu
\thinmuskip=-2mu
\thickmuskip=-2mu
\nulldelimiterspace=-1pt \begin{array}{ccc|ccc}
1+D       & 1+D  & 1+D+D^2 & 1 & 0 & 0\\
D         & D    &  1+D^2  & 0 & 1 & 0\\
1+D       & D    & D^2     & 0 & 0 & 1
\end{array}}
\right]\hspace{-1mm}, \label{eq:pseudo_conv_inverse3}\\
\left[\begin{array}{c|c}I&\begin{array}{c}G(D)\\ \text{ISF}\end{array}\end{array}\right] &= \left[
{\medmuskip=-2mu
\thinmuskip=-2mu
\thickmuskip=-2mu\begin{array}{ccc|ccc}
1       & 0 & 0   & D       & D     & 1\\
0       & 1   & 0 & 1+D+D^2 & 1+D^2 & 1\\
0       & 0   &1  & D       & 1+D   & 0
\end{array}}
\right]\hspace{-1mm}.\label{eq:pseudo_conv_inverse4}
\end{align}
The inverse syndrome former $\textrm{ISF}$ is given by
\begin{align}
\textrm{ISF}(D) = \left[ \begin{array}{C{1cm}C{1.5cm}C{1cm}} $D$ & $1+D$ & $0$ \end{array} \right].
\end{align}
The product of $\textrm{ISF}(D)$ and $H(D)$ is exactly 1. We can express this as:
\begin{align}
\textrm{ISF}(D^i) \times H(D^j) = \delta_{ij}.
\end{align}
For an arbitrary syndrome we can use the $\textrm{ISF}$ to generate an error pattern which satisfies the syndrome.

\section{Syndrome and Error pattern Calculations}
\label{eScalc}

Here we show how to calculate the syndrome, $\vec S$, and initial error pattern, $\vec{e}^{\,0}$, in the examples given in section~\ref{sec:convolutional_trellis_construction}. The results appearing in \cref{eq:blue_path} are calculated using $H$, $\vec \varepsilon$, and the ISF. The complete list of stabilizers is given by translations of the seed stabilizer by $D^j$. The index, $j$, gives the $j^\textrm{th}$ row of the parity check matrix.  We have, for all $j\in \{1, ..., \tau \}$,
\begin{align*}
\mathbf{H}_j &= \left[\begin{array}{C{2cm}C{2cm}C{1cm}}
${\color{Orange}{1}}+{\color{Green}{D}}+{\color{Purple}{D^2}}$ & ${\color{Cyan}{1}}+{\color{magenta}{D^2}}$ & ${\color{Yellow}{1}}$ 
\end{array} \right]D^j,\\
\mathbf{ISF} &= \left[\begin{array}{C{2cm}C{2cm}C{1cm}}
$D$ & $D$ & $0$ 
\end{array} \right].
\end{align*}
For reference, we also write $\mathbf{H}$ in the less compact but more direct  notation  of \cref{eq:uncompactH}, using the colours to match terms above with locations of 1's in the first row below; subsequent rows are translations of the top row 
\begin{align}
\mathbf{H} \!\!=\! \left[ \begin{array} {c cc ccc ccc}
\hdots& 000& {\color{Orange}{1}} {\color{Cyan}{1}} {\color{Yellow}{1}}        & {\color{Green}{1}}00 &  {\color{Purple}{1}}{\color{magenta}{1}}0 & 000 &   \hdots   & &\\
&\hdots& 000& 111        &100 & 110 & 000 &   \hdots   & \\
&&\hdots& 000& 111        &100 & 110 & 000 &   \hdots   
\end{array}\right]_{{n_z}\tau\times n\tau},\nonumber\end{align}

\tmsB{For the blue path in the example of \cref{eq:blue_path}, the error pattern, expressed in delay notation is}
\begin{equation}
\vec \varepsilon = \left[\begin{array}{C{2cm}C{2cm}C{1cm}}
$D$ & $0$ & $0$ 
\end{array} \right]D^t,
\end{equation}
The $j^\textrm{th}$ element of the syndrome is then given by
\begin{align*}
S_j&=\mathbf{H}_j\cdot\vec \varepsilon\transpose(\widetilde{D}),\\
&= D^{j}(1+D+D^2)\widetilde{D}^{t+1},\\
        &= \delta_{j,t+1} + \delta_{j+1,t+1} + \delta_{j+2,t+1},\\
         &= \delta_{j,t+1} + \delta_{j,t} + \delta_{j,t-1},
\end{align*}
\tmsB{which corresponds to the syndrome, $\vec S$, listed in \cref{eq:blue_path}, with 1's in the syndrome at frames $t-1,t$ and $t+1$, and zero everywhere else.}

In delay notation, 
\begin{equation}
\vec S=\sum_j S_j D^j=D^{t+1}+D^t+D^{t-1},
\end{equation}
and then using the ISF and the syndrome we calculate the initial error pattern
\begin{align*}
\vec{e}^{\,0} &= \mathbf{ISF}\cdot\vec S, \\
&=  \left[\begin{array}{ccc} 
D^t+D^{t+1}+D^{t+2} \quad{}& D^t+D^{t+1}+D^{t+2}\quad{} & 0 
\end{array} \right],
\end{align*}
which is expressed as the string of bits $...110 \hspace{2mm}110 \hspace{2mm}110...$ in \cref{eq:blue_path}, with the first triplet belonging to frame $t$. 

The example using the red path, as shown in \cref{eq:red_path}, uses the same code. As such the terms for $H$ and ISF are the same as the previous example. The error pattern used in this example is
\begin{align*}
\vec \varepsilon &= \left[\begin{array}{C{2cm}C{2cm}C{1cm}}
$D$ & $D$ & $0$ 
\end{array} \right]D^t.
\end{align*}
This generates the $j^\textrm{th}$ element of the syndrome
\begin{align*}
{S}_j &= D^{j}(1+D+D^2)\widetilde{D}^{t+1}+D^{j}(1+D^2)\widetilde{D}^{t+1},\\
&= D^{j+1} \widetilde{D}^{t+1},\\
&= \delta_{j+1,t+1}, \\
&= \delta_{j,t}.
\end{align*}
In delay notation, 
$
\vec S=D^t
$,
and then using the ISF and the syndrome we calculate the initial error pattern
This agrees with the syndrome in \cref{eq:red_path}
The initial error pattern is given by 
\begin{align*}
\vec{e}^{\,0} &= \mathbf{ISF}\cdot\vec S, \\
&=  \left[\begin{array}{C{2cm}C{2cm}C{1cm}}
$D^{t+1}$ & $D^{t+1}$ & $0$ 
\end{array} \right],
\end{align*}
which is expressed as the string of bits $110$  in frame $t+1$ of in \cref{eq:red_path}.

\section{Schedule for constructing the T25 cluster code}
\label{T25schedule}

\begin{figure}[t]
\begin{center}
\includegraphics[width=\columnwidth]{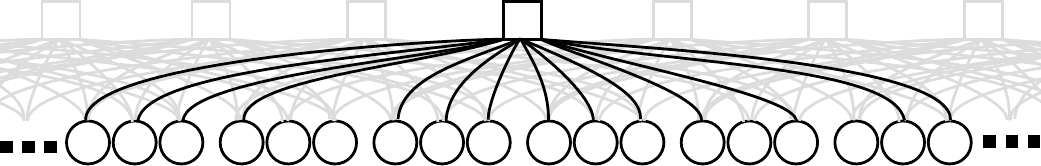}
\caption{The cluster state resource for the $[[n,n/3,5]]$  C5 code.  \tmsB{The ancilla (square) shares cluster bonds (lines) with the code qubits (circles).}  A single stabilizer of weight 14 is illustrated in black, and all other stabilizers are a translation of this seed. Construction of this cluster state requires assigning a schedule for each bond in the stabilizer. The same scheduling can be used in parallel with the other stabilizers, such that the total number of time steps required to generate the entire cluster is 14.} \label{fig:C5_schematic}
\end{center}
\end{figure}

The C5 code is a self-dual rate $r = \frac{1}{3}$ code with stabilizers generated by
\begin{align}
H \!= \!\left[\! \begin{array}{C{2cm}C{19mm}C{19mm}}
$1+D+D^2$ & $1+D^2+D^3$  & $1+D^2+ D^3$\\
$+D^3+D^4$ & $+D^5$ & $+ D^4+D^5$
\end{array}\!\right]\!.
\end{align}
The weight of this stabilizer is 14, and as such a clusterized code can be constructed in 14 time-steps. \fig{fig:C5_schematic} depicts the form of the cluster state used to construct the clusterized C5 code. One possible scheduling for gate operations is recorded below using $\Lambda(a_{i,j},c_{i',j'},T_m)$ for gate operations, where $i$ refers to the $i$th ancilla qubit within a frame and $j$ refers to the $j$th frame of the code and $T_m$ is a time index. The terms $a$ and $c$ denote ancilla and code qubits respectively. The C5 code is a rate $r = \frac{1}{3}$ code and as such there are 3 code qubits and 1 ancilla qubit per frame. A possible scheduling for gate operations is given by
\begin{align}
&\text{Order C3} = \nonumber\\
&\begin{array}{ll}
	\Lambda_{T_1}(a_{1,t},c_{1,t}), & \Lambda_{T_2}(a_{1,t},c_{3,t+4}),\\ \Lambda_{T_3}(a_{1,t},c_{1,t+1}, &\Lambda_{T_4}(a_{1,t},c_{3,t+3}), \\	\Lambda_{T_5}(a_{1,t},c_{3,t+2}), &\Lambda_{T_6}(a_{1,t},c_{1,t+3}),\\
\Lambda_{T_7}(a_{1,t},c_{1,t+2},&\Lambda_{T_8}(a_{1,t},c_{2,t}),\\ \Lambda_{T_9}(a_{1,t},c_{2,t+3}),& \Lambda_{T_{10}}(a_{1,t},c_{3,t+5}), \\ \Lambda_{T_{11}}(a_{1,t},c_{3,t}), & \Lambda_{T_{12}}(a_{1,t},c_{2,t+2}), \\
\Lambda_{T_{13}}(a_{1,t},c_{2,t+5}), & \Lambda_{T_{14}}(a_{1,t},c_{1,t+4}). 
\end{array}
\end{align}

Encoding these stabilizers with another C5 convolutional code produces a turbo code whose outer stabilizers are weight $7 \times 14 = 98$, where $7$ is the weight of the generators, $G_{C5}$. The weight of these stabilizers is very high. 

To reduce the effective weight of these outer stabilizers it is possible to make a choice of interleaver such that the weight of these stabilizers is greatly reduced by taking operator products with inner stabilizers (whose form is identical to the scheduling outlined for the C3 code above). One such choice of interleaver is to generate independent inner encodings for set of qubits produced by the outer encoder, as determined by their position within a frame. For example encode all of the qubits which are in the first position in every frame, then encode all the qubits which are in the second position etc. This produces an inner encoding where logical qubits are more closely correlated than a random interleaver. The benefit is that an inner stabilizer can be generated with a weight of only 26, as compared to 98. 

One possible scheduling for a weight 26 stabilizer produced by this choice of interleaver is given by
\begin{align}
&\text{Order T25}(1) = \nonumber\\
&\begin{array}{ll}
	\Lambda_{T_1}(A_{1,t},c_{2,t}), & \Lambda_{T_4}(A_{1,t},c_{2,t+5}),\\   	\Lambda_{T_7}(A_{1,t},c_{2,t+1}), &\Lambda_{T_{10}}(A_{1,t},c_{2,t+3}),\\
	\Lambda_{T_{13}}(A_{1,t},c_{3,t+2}), &\Lambda_{T_{16}}(A_{1,t},c_{3,t+5}),\\
\Lambda_{T_{19}}(A_{1,t},c_{3,t}),& \Lambda_{T_{20}}(A_{1,t},c_{3,t+3}),\\
\Lambda_{T_{23}}(A_{1,t},c_{2,t+2}).&
\end{array}\\
&\text{Order T25}(2) = \nonumber\\ 
&\begin{array}{ll}
\Lambda_{T_{2}}(A_{1,t},c_{2,t'}), & \Lambda_{T_{5}}(A_{1,t},c_{1,t'+5}), \\
\Lambda_{T_{8}}(A_{1,t},c_{3,t'+1}), &\Lambda_{T_{11}}(A_{1,t},c_{2,t'+3}),\\
\Lambda_{T_{14}}(A_{1,t},c_{2,t'+5}), &\Lambda_{T_{17}}(A_{1,t},c_{3,t'}),\\
\Lambda_{T_{21}}(A_{1,t},c_{1,t'+4}),&\Lambda_{T_{24}}(A_{1,t},c_{2,t'+2}). 
\end{array}\\
&\text{Order T25}(3) = \nonumber\\
&\begin{array}{ll}
	\Lambda_{T_{3}}(A_{1,t},c_{2,t''}), & \Lambda_{T_{6}}(A_{1,t},c_{2,t''+6}),\\
\Lambda_{T_{9}}(A_{1,t},c_{3,t''+1}), &\Lambda_{T_{12}}(A_{1,t},c_{3,t''+2}),\\
\Lambda_{T_{15}}(A_{1,t},c_{3,t''+6}), &\Lambda_{T_{18}}(A_{1,t},c_{3,t''}), \\
	\Lambda_{T_{22}}(A_{1,t},c_{2,t''+3}),& \Lambda_{T_{25}}(A_{1,t},c_{1,t''+2}),\\ \Lambda_{T_{26}}(A_{1,t},c_{1,t''+3}). &
\end{array}
\end{align}
The scheduling has been split into three components, (1), (2) and (3), referring to each of the outputs bitstreams from the outer encoder. The frames indices $t$, $t'$ and $t''$ refer to frames which are removed from each other, such that each code qubit in the scheduling is uniquely identified by the scheme. Finally we have substituted $a$ with $A$ to refer to the outer ancilla qubits.

\bibliography{bibliography}

\begin{thebibliography}{49}
\expandafter\ifx\csname natexlab\endcsname\relax\def\natexlab#1{#1}\fi
\expandafter\ifx\csname bibnamefont\endcsname\relax
  \def\bibnamefont#1{#1}\fi
\expandafter\ifx\csname bibfnamefont\endcsname\relax
  \def\bibfnamefont#1{#1}\fi
\expandafter\ifx\csname citenamefont\endcsname\relax
  \def\citenamefont#1{#1}\fi
\expandafter\ifx\csname url\endcsname\relax
  \def\url#1{\texttt{#1}}\fi
\expandafter\ifx\csname urlprefix\endcsname\relax\def\urlprefix{URL }\fi
\providecommand{\bibinfo}[2]{#2}
\providecommand{\eprint}[2][]{\url{#2}}

\bibitem[{\citenamefont{Shor}(1995)}]{shor_1995}
\bibinfo{author}{\bibfnamefont{P.}~\bibnamefont{Shor}}, \bibinfo{journal}{Phys.
  Rev. A.} \textbf{\bibinfo{volume}{52-4}}, \bibinfo{pages}{R2493}
  (\bibinfo{year}{1995}).

\bibitem[{\citenamefont{Steane}(1996)}]{steane_1996}
\bibinfo{author}{\bibfnamefont{A.~M.} \bibnamefont{Steane}},
  \bibinfo{journal}{Phys. Rev. Lett.} \textbf{\bibinfo{volume}{77}},
  \bibinfo{pages}{793} (\bibinfo{year}{1996}).

\bibitem[{\citenamefont{Raussendorf and Briegel}(2001)}]{raussendorf_2001}
\bibinfo{author}{\bibfnamefont{R.}~\bibnamefont{Raussendorf}} \bibnamefont{and}
  \bibinfo{author}{\bibfnamefont{H.~J.} \bibnamefont{Briegel}},
  \bibinfo{journal}{Phys. Rev. Lett.} \textbf{\bibinfo{volume}{86}},
  \bibinfo{pages}{5188} (\bibinfo{year}{2001}).

\bibitem[{\citenamefont{Raussendorf et~al.}(2003)\citenamefont{Raussendorf,
  Browne, and Briegel}}]{raussendorf_2003}
\bibinfo{author}{\bibfnamefont{R.}~\bibnamefont{Raussendorf}},
  \bibinfo{author}{\bibfnamefont{D.~E.} \bibnamefont{Browne}},
  \bibnamefont{and} \bibinfo{author}{\bibfnamefont{H.~J.}
  \bibnamefont{Briegel}}, \bibinfo{journal}{Phys. Rev. A.}
  \textbf{\bibinfo{volume}{68}}, \bibinfo{pages}{022312}
  (\bibinfo{year}{2003}).

\bibitem[{\citenamefont{Raussendorf et~al.}(2005)\citenamefont{Raussendorf,
  Bravyi, and Harrington}}]{raussendorf_2005}
\bibinfo{author}{\bibfnamefont{R.}~\bibnamefont{Raussendorf}},
  \bibinfo{author}{\bibfnamefont{S.}~\bibnamefont{Bravyi}}, \bibnamefont{and}
  \bibinfo{author}{\bibfnamefont{J.}~\bibnamefont{Harrington}},
  \bibinfo{journal}{Phys. Rev. A.} \textbf{\bibinfo{volume}{71}},
  \bibinfo{pages}{062313} (\bibinfo{year}{2005}).

\bibitem[{\citenamefont{Raussendorf et~al.}(2006)\citenamefont{Raussendorf,
  Harrington, and Goyal}}]{raussendorf_2006}
\bibinfo{author}{\bibfnamefont{R.}~\bibnamefont{Raussendorf}},
  \bibinfo{author}{\bibfnamefont{J.}~\bibnamefont{Harrington}},
  \bibnamefont{and} \bibinfo{author}{\bibfnamefont{K.}~\bibnamefont{Goyal}},
  \bibinfo{journal}{Annals of Physics} \textbf{\bibinfo{volume}{321-9}},
  \bibinfo{pages}{2242} (\bibinfo{year}{2006}).

\bibitem[{\citenamefont{Raussendorf and Harrington}(2007)}]{raussendorf_2007a}
\bibinfo{author}{\bibfnamefont{R.}~\bibnamefont{Raussendorf}} \bibnamefont{and}
  \bibinfo{author}{\bibfnamefont{J.}~\bibnamefont{Harrington}},
  \bibinfo{journal}{Phys. Rev. Letters} \textbf{\bibinfo{volume}{98}},
  \bibinfo{pages}{190504} (\bibinfo{year}{2007}).

\bibitem[{\citenamefont{Raussendorf et~al.}(2007)\citenamefont{Raussendorf,
  Harrington, and Goyal}}]{raussendorf_2007b}
\bibinfo{author}{\bibfnamefont{R.}~\bibnamefont{Raussendorf}},
  \bibinfo{author}{\bibfnamefont{J.}~\bibnamefont{Harrington}},
  \bibnamefont{and} \bibinfo{author}{\bibfnamefont{K.}~\bibnamefont{Goyal}},
  \bibinfo{journal}{New Journ. Phys.} \textbf{\bibinfo{volume}{9}},
  \bibinfo{pages}{199} (\bibinfo{year}{2007}).

\bibitem[{\citenamefont{Kitaev}(1997)}]{kitaev_1997}
\bibinfo{author}{\bibfnamefont{A.}~\bibnamefont{Kitaev}},
  \bibinfo{journal}{Annals of Physics} \textbf{\bibinfo{volume}{303-1}},
  \bibinfo{pages}{2} (\bibinfo{year}{1997}).

\bibitem[{\citenamefont{Dennis et~al.}(2002)\citenamefont{Dennis, Kitaev,
  Landahl, and Preskill}}]{dennis_2002}
\bibinfo{author}{\bibfnamefont{E.}~\bibnamefont{Dennis}},
  \bibinfo{author}{\bibfnamefont{A.}~\bibnamefont{Kitaev}},
  \bibinfo{author}{\bibfnamefont{A.}~\bibnamefont{Landahl}}, \bibnamefont{and}
  \bibinfo{author}{\bibfnamefont{J.}~\bibnamefont{Preskill}},
  \bibinfo{journal}{Journ. Math. Phys.} \textbf{\bibinfo{volume}{43-9}},
  \bibinfo{pages}{4452} (\bibinfo{year}{2002}).

\bibitem[{\citenamefont{Fowler et~al.}(2012)\citenamefont{Fowler, Mariantoni,
  Martinis, and Cleland}}]{PhysRevA.86.032324}
\bibinfo{author}{\bibfnamefont{A.~G.} \bibnamefont{Fowler}},
  \bibinfo{author}{\bibfnamefont{M.}~\bibnamefont{Mariantoni}},
  \bibinfo{author}{\bibfnamefont{J.~M.} \bibnamefont{Martinis}},
  \bibnamefont{and} \bibinfo{author}{\bibfnamefont{A.~N.}
  \bibnamefont{Cleland}}, \bibinfo{journal}{Phys. Rev. A}
  \textbf{\bibinfo{volume}{86}}, \bibinfo{pages}{032324}
  (\bibinfo{year}{2012}),
  \urlprefix\url{https://link.aps.org/doi/10.1103/PhysRevA.86.032324}.

\bibitem[{\citenamefont{Martinis et~al.}(2014)\citenamefont{Martinis, Barends,
  Kelly, Veitia, Sank, Jeffrey, White, Mutus, Fowler, Campbell
  et~al.}}]{martinis_2014}
\bibinfo{author}{\bibfnamefont{J.}~\bibnamefont{Martinis}},
  \bibinfo{author}{\bibfnamefont{R.}~\bibnamefont{Barends}},
  \bibinfo{author}{\bibfnamefont{J.}~\bibnamefont{Kelly}},
  \bibinfo{author}{\bibfnamefont{A.}~\bibnamefont{Veitia}},
  \bibinfo{author}{\bibfnamefont{D.}~\bibnamefont{Sank}},
  \bibinfo{author}{\bibfnamefont{E.}~\bibnamefont{Jeffrey}},
  \bibinfo{author}{\bibfnamefont{T.}~\bibnamefont{White}},
  \bibinfo{author}{\bibfnamefont{J.}~\bibnamefont{Mutus}},
  \bibinfo{author}{\bibfnamefont{A.}~\bibnamefont{Fowler}},
  \bibinfo{author}{\bibfnamefont{B.}~\bibnamefont{Campbell}},
  \bibnamefont{et~al.}, \bibinfo{journal}{Nature, 508}  (\bibinfo{year}{2014}).

\bibitem[{\citenamefont{Rist{\`e} et~al.}(2015)\citenamefont{Rist{\`e},
  Poletto, Huang, Bruno, Vesterinen, Saira, and DiCarlo}}]{Riste:2015aa}
\bibinfo{author}{\bibfnamefont{D.}~\bibnamefont{Rist{\`e}}},
  \bibinfo{author}{\bibfnamefont{S.}~\bibnamefont{Poletto}},
  \bibinfo{author}{\bibfnamefont{M.~Z.} \bibnamefont{Huang}},
  \bibinfo{author}{\bibfnamefont{A.}~\bibnamefont{Bruno}},
  \bibinfo{author}{\bibfnamefont{V.}~\bibnamefont{Vesterinen}},
  \bibinfo{author}{\bibfnamefont{O.~P.} \bibnamefont{Saira}}, \bibnamefont{and}
  \bibinfo{author}{\bibfnamefont{L.}~\bibnamefont{DiCarlo}},
  \bibinfo{journal}{Nature Communications} \textbf{\bibinfo{volume}{6}},
  \bibinfo{pages}{6983 EP } (\bibinfo{year}{2015}),
  \urlprefix\url{http://dx.doi.org/10.1038/ncomms7983}.

\bibitem[{\citenamefont{Gambetta et~al.}(2017)\citenamefont{Gambetta, Chow, and
  Steffen}}]{Gambetta:2017aa}
\bibinfo{author}{\bibfnamefont{J.~M.} \bibnamefont{Gambetta}},
  \bibinfo{author}{\bibfnamefont{J.~M.} \bibnamefont{Chow}}, \bibnamefont{and}
  \bibinfo{author}{\bibfnamefont{M.}~\bibnamefont{Steffen}},
  \bibinfo{journal}{npj Quantum Information} \textbf{\bibinfo{volume}{3}},
  \bibinfo{pages}{2} (\bibinfo{year}{2017}).

\bibitem[{\citenamefont{Stace and Barrett}(2010)}]{stace_2010}
\bibinfo{author}{\bibfnamefont{T.~M.} \bibnamefont{Stace}} \bibnamefont{and}
  \bibinfo{author}{\bibfnamefont{S.~D.} \bibnamefont{Barrett}},
  \bibinfo{journal}{Phys. Rev. A.} \textbf{\bibinfo{volume}{81}},
  \bibinfo{pages}{022317} (\bibinfo{year}{2010}).

\bibitem[{\citenamefont{Barrett and Stace}(2010)}]{barrett_2010}
\bibinfo{author}{\bibfnamefont{S.~D.} \bibnamefont{Barrett}} \bibnamefont{and}
  \bibinfo{author}{\bibfnamefont{T.~M.} \bibnamefont{Stace}},
  \bibinfo{journal}{Phys. Rev. Lett.} \textbf{\bibinfo{volume}{105}},
  \bibinfo{pages}{200502} (\bibinfo{year}{2010}),
  \urlprefix\url{https://link.aps.org/doi/10.1103/PhysRevLett.105.200502}.

\bibitem[{\citenamefont{Duclos-Cianci and Poulin}(2010)}]{duclos_2010}
\bibinfo{author}{\bibfnamefont{G.}~\bibnamefont{Duclos-Cianci}}
  \bibnamefont{and} \bibinfo{author}{\bibfnamefont{D.}~\bibnamefont{Poulin}},
  \bibinfo{journal}{Phys. Rev. Lett.} \textbf{\bibinfo{volume}{104}},
  \bibinfo{pages}{050504} (\bibinfo{year}{2010}),
  \urlprefix\url{https://link.aps.org/doi/10.1103/PhysRevLett.104.050504}.

\bibitem[{\citenamefont{Rudolph}(2017)}]{rudolph2017optimistic}
\bibinfo{author}{\bibfnamefont{T.}~\bibnamefont{Rudolph}},
  \bibinfo{journal}{APL Photonics} \textbf{\bibinfo{volume}{2}},
  \bibinfo{pages}{030901} (\bibinfo{year}{2017}).

\bibitem[{\citenamefont{Bravyi and Kitaev}(2005)}]{bravyi_2005}
\bibinfo{author}{\bibfnamefont{S.}~\bibnamefont{Bravyi}} \bibnamefont{and}
  \bibinfo{author}{\bibfnamefont{A.}~\bibnamefont{Kitaev}},
  \bibinfo{journal}{Physical Review A} \textbf{\bibinfo{volume}{71}},
  \bibinfo{pages}{022316} (\bibinfo{year}{2005}).

\bibitem[{\citenamefont{Bolt et~al.}(2016)\citenamefont{Bolt, Duclos-Cianci,
  Poulin, and Stace}}]{bolt_2016}
\bibinfo{author}{\bibfnamefont{A.}~\bibnamefont{Bolt}},
  \bibinfo{author}{\bibfnamefont{G.}~\bibnamefont{Duclos-Cianci}},
  \bibinfo{author}{\bibfnamefont{D.}~\bibnamefont{Poulin}}, \bibnamefont{and}
  \bibinfo{author}{\bibfnamefont{T.~M.} \bibnamefont{Stace}},
  \bibinfo{journal}{Phys. Rev. Lett.} \textbf{\bibinfo{volume}{117}},
  \bibinfo{pages}{070501} (\bibinfo{year}{2016}).

\bibitem[{\citenamefont{Berrou et~al.}(1993)\citenamefont{Berrou, Ecole,
  Glavieux, and Thitmajshima}}]{berrou_1993}
\bibinfo{author}{\bibfnamefont{C.}~\bibnamefont{Berrou}},
  \bibinfo{author}{\bibfnamefont{N.}~\bibnamefont{Ecole}},
  \bibinfo{author}{\bibfnamefont{A.}~\bibnamefont{Glavieux}}, \bibnamefont{and}
  \bibinfo{author}{\bibfnamefont{P.}~\bibnamefont{Thitmajshima}}, in
  \emph{\bibinfo{booktitle}{Communications. ICC '93 Geneva. Technical Program,
  Conference Record, IEEE International Conference on (vol 2)}}
  (\bibinfo{year}{1993}).

\bibitem[{\citenamefont{McEliece et~al.}(1998)\citenamefont{McEliece, MacKay,
  and Cheng}}]{mceliece_1998}
\bibinfo{author}{\bibfnamefont{R.}~\bibnamefont{McEliece}},
  \bibinfo{author}{\bibfnamefont{D.}~\bibnamefont{MacKay}}, \bibnamefont{and}
  \bibinfo{author}{\bibfnamefont{J.}~\bibnamefont{Cheng}},
  \bibinfo{journal}{IEEE Journal on Selected Areas in Communications}
  \textbf{\bibinfo{volume}{16}}, \bibinfo{pages}{140} (\bibinfo{year}{1998}).

\bibitem[{\citenamefont{Poulin and Tillich}(2009)}]{poulin_2009}
\bibinfo{author}{\bibfnamefont{D.}~\bibnamefont{Poulin}} \bibnamefont{and}
  \bibinfo{author}{\bibfnamefont{H.}~\bibnamefont{Tillich},
  \bibfnamefont{J.-P.~Ollivier}}, \bibinfo{journal}{IEEE Transactions on
  Information Theory} \textbf{\bibinfo{volume}{55 No. 6}},
  \bibinfo{pages}{2776} (\bibinfo{year}{2009}).

\bibitem[{\citenamefont{Utby}(2006)}]{utby_2006}
\bibinfo{author}{\bibfnamefont{H.}~\bibnamefont{Utby}}, Master's thesis,
  \bibinfo{school}{University of Bergen} (\bibinfo{year}{2006}).

\bibitem[{\citenamefont{Briegel et~al.}(2009)\citenamefont{Briegel, Browne,
  Dur, Raussendorf, and Nest}}]{briegel_2009}
\bibinfo{author}{\bibfnamefont{H.}~\bibnamefont{Briegel}},
  \bibinfo{author}{\bibfnamefont{D.}~\bibnamefont{Browne}},
  \bibinfo{author}{\bibfnamefont{W.}~\bibnamefont{Dur}},
  \bibinfo{author}{\bibfnamefont{R.}~\bibnamefont{Raussendorf}},
  \bibnamefont{and} \bibinfo{author}{\bibfnamefont{V.~d.} \bibnamefont{Nest}},
  \bibinfo{journal}{Nature Physics} \textbf{\bibinfo{volume}{5}},
  \bibinfo{pages}{19} (\bibinfo{year}{2009}).

\bibitem[{\citenamefont{Tanner}(1981)}]{tanner_1981}
\bibinfo{author}{\bibfnamefont{R.}~\bibnamefont{Tanner}},
  \bibinfo{journal}{IEEE Trans. Inform. Theory} \textbf{\bibinfo{volume}{27-5}}
  (\bibinfo{year}{1981}).

\bibitem[{\citenamefont{Ollivier and Tillich}(2003)}]{ollivier_2003}
\bibinfo{author}{\bibfnamefont{H.}~\bibnamefont{Ollivier}} \bibnamefont{and}
  \bibinfo{author}{\bibfnamefont{J.-P.} \bibnamefont{Tillich}},
  \bibinfo{journal}{Phys. Rev. Lett.} \textbf{\bibinfo{volume}{91}},
  \bibinfo{pages}{177902} (\bibinfo{year}{2003}).

\bibitem[{\citenamefont{Kitaev}(2003)}]{kitaev_2003}
\bibinfo{author}{\bibfnamefont{A.}~\bibnamefont{Kitaev}},
  \bibinfo{journal}{Annals of Physics} \textbf{\bibinfo{volume}{303-1}},
  \bibinfo{pages}{2} (\bibinfo{year}{2003}).

\bibitem[{\citenamefont{Tillich and Zemor}(2009)}]{tillich_2009}
\bibinfo{author}{\bibfnamefont{J.-P.} \bibnamefont{Tillich}} \bibnamefont{and}
  \bibinfo{author}{\bibfnamefont{G.}~\bibnamefont{Zemor}}, in
  \emph{\bibinfo{booktitle}{Information Theory, IEEE International Symposium
  on}} (\bibinfo{year}{2009}).

\bibitem[{\citenamefont{Poulin and Chung}(2008)}]{poulin_2008}
\bibinfo{author}{\bibfnamefont{D.}~\bibnamefont{Poulin}} \bibnamefont{and}
  \bibinfo{author}{\bibfnamefont{Y.}~\bibnamefont{Chung}},
  \bibinfo{journal}{Quantum Information \& Computation, 8-10}
  (\bibinfo{year}{2008}).

\bibitem[{\citenamefont{Forney}(1965)}]{forney_1965}
\bibinfo{author}{\bibfnamefont{D.}~\bibnamefont{Forney}}, Ph.D. thesis,
  \bibinfo{school}{Massachusetts Institute of Technology}
  (\bibinfo{year}{1965}).

\bibitem[{\citenamefont{Sadjadpour}(2000)}]{sadjadpour_2000}
\bibinfo{author}{\bibfnamefont{H.}~\bibnamefont{Sadjadpour}},
  \bibinfo{journal}{Proc. SPIE 4045, Digital Wireless Communication 2, 73}
  (\bibinfo{year}{2000}).

\bibitem[{\citenamefont{Viterbi}(1967)}]{viterbi_1967}
\bibinfo{author}{\bibfnamefont{A.}~\bibnamefont{Viterbi}},
  \bibinfo{journal}{IEEE Trans. Inf. Theory 13}  (\bibinfo{year}{1967}).

\bibitem[{\citenamefont{Bahl et~al.}(1974)\citenamefont{Bahl, Cocke, Jelinek,
  and Raviv}}]{bahl_1974}
\bibinfo{author}{\bibfnamefont{L.}~\bibnamefont{Bahl}},
  \bibinfo{author}{\bibfnamefont{J.}~\bibnamefont{Cocke}},
  \bibinfo{author}{\bibfnamefont{F.}~\bibnamefont{Jelinek}}, \bibnamefont{and}
  \bibinfo{author}{\bibfnamefont{J.}~\bibnamefont{Raviv}},
  \bibinfo{journal}{IEEE Trans. Inf. Theory} \textbf{\bibinfo{volume}{20}},
  \bibinfo{pages}{284} (\bibinfo{year}{1974}).

\bibitem[{\citenamefont{Benedetto et~al.}(1996)\citenamefont{Benedetto,
  Divsalar, Montorsi, and Pollara}}]{benedetto_1996}
\bibinfo{author}{\bibfnamefont{S.}~\bibnamefont{Benedetto}},
  \bibinfo{author}{\bibfnamefont{D.}~\bibnamefont{Divsalar}},
  \bibinfo{author}{\bibfnamefont{G.}~\bibnamefont{Montorsi}}, \bibnamefont{and}
  \bibinfo{author}{\bibfnamefont{F.}~\bibnamefont{Pollara}},
  \bibinfo{journal}{TDA Progress Report. 42-124.}  (\bibinfo{year}{1996}).

\bibitem[{\citenamefont{MacKay and Neal}(1997)}]{mackay_1997}
\bibinfo{author}{\bibfnamefont{D.}~\bibnamefont{MacKay}} \bibnamefont{and}
  \bibinfo{author}{\bibfnamefont{R.}~\bibnamefont{Neal}},
  \bibinfo{journal}{Electronic Lett. 33}  (\bibinfo{year}{1997}).

\bibitem[{\citenamefont{Mackay et~al.}(2004)\citenamefont{Mackay, Mitchison,
  and McFadden}}]{mackay_2004}
\bibinfo{author}{\bibfnamefont{D.}~\bibnamefont{Mackay}},
  \bibinfo{author}{\bibfnamefont{G.}~\bibnamefont{Mitchison}},
  \bibnamefont{and} \bibinfo{author}{\bibfnamefont{P.}~\bibnamefont{McFadden}},
  \bibinfo{journal}{IEEE Transactions on information theory}
  \textbf{\bibinfo{volume}{50-10}}, \bibinfo{pages}{2315}
  (\bibinfo{year}{2004}).

\bibitem[{\citenamefont{Pearl}(1985)}]{pearl_1985}
\bibinfo{author}{\bibfnamefont{J.}~\bibnamefont{Pearl}}, in
  \emph{\bibinfo{booktitle}{Proceedings of the 7th Conference of the Cognitive
  Science Society, University of California.}} (\bibinfo{year}{1985}).

\bibitem[{\citenamefont{Li et~al.}(2010)\citenamefont{Li, Barrett, Stace, and
  Benjamin}}]{li_2010}
\bibinfo{author}{\bibfnamefont{Y.}~\bibnamefont{Li}},
  \bibinfo{author}{\bibfnamefont{S.~D.} \bibnamefont{Barrett}},
  \bibinfo{author}{\bibfnamefont{T.~M.} \bibnamefont{Stace}}, \bibnamefont{and}
  \bibinfo{author}{\bibfnamefont{S.~C.} \bibnamefont{Benjamin}},
  \bibinfo{journal}{Phys. Rev. Lett} \textbf{\bibinfo{volume}{105}},
  \bibinfo{pages}{250502} (\bibinfo{year}{2010}).

\bibitem[{\citenamefont{Duan et~al.}(2001)\citenamefont{Duan, Lukin, Cirac, and
  Zoller}}]{duan_2001}
\bibinfo{author}{\bibfnamefont{L.-M.} \bibnamefont{Duan}},
  \bibinfo{author}{\bibfnamefont{M.}~\bibnamefont{Lukin}},
  \bibinfo{author}{\bibfnamefont{J.}~\bibnamefont{Cirac}}, \bibnamefont{and}
  \bibinfo{author}{\bibfnamefont{P.}~\bibnamefont{Zoller}},
  \bibinfo{journal}{Nature} \textbf{\bibinfo{volume}{414}},
  \bibinfo{pages}{413} (\bibinfo{year}{2001}).

\bibitem[{\citenamefont{Benjamin et~al.}(2005)\citenamefont{Benjamin, Eisert,
  and Stace}}]{1367-2630-7-1-194}
\bibinfo{author}{\bibfnamefont{S.~C.} \bibnamefont{Benjamin}},
  \bibinfo{author}{\bibfnamefont{J.}~\bibnamefont{Eisert}}, \bibnamefont{and}
  \bibinfo{author}{\bibfnamefont{T.~M.} \bibnamefont{Stace}},
  \bibinfo{journal}{New Journal of Physics} \textbf{\bibinfo{volume}{7}},
  \bibinfo{pages}{194} (\bibinfo{year}{2005}),
  \urlprefix\url{http://stacks.iop.org/1367-2630/7/i=1/a=194}.

\bibitem[{\citenamefont{Barrett et~al.}(2010)\citenamefont{Barrett, Rohde, and
  Stace}}]{1367-2630-12-9-093032}
\bibinfo{author}{\bibfnamefont{S.~D.} \bibnamefont{Barrett}},
  \bibinfo{author}{\bibfnamefont{P.~P.} \bibnamefont{Rohde}}, \bibnamefont{and}
  \bibinfo{author}{\bibfnamefont{T.~M.} \bibnamefont{Stace}},
  \bibinfo{journal}{New Journal of Physics} \textbf{\bibinfo{volume}{12}},
  \bibinfo{pages}{093032} (\bibinfo{year}{2010}),
  \urlprefix\url{http://stacks.iop.org/1367-2630/12/i=9/a=093032}.

\bibitem[{\citenamefont{Satoh et~al.}(2016)\citenamefont{Satoh, Ishizaki,
  Nagayama, and Van~Meter}}]{satoh_2016}
\bibinfo{author}{\bibfnamefont{T.}~\bibnamefont{Satoh}},
  \bibinfo{author}{\bibfnamefont{K.}~\bibnamefont{Ishizaki}},
  \bibinfo{author}{\bibfnamefont{S.}~\bibnamefont{Nagayama}}, \bibnamefont{and}
  \bibinfo{author}{\bibfnamefont{R.}~\bibnamefont{Van~Meter}},
  \bibinfo{journal}{Phys. Rev. A.} \textbf{\bibinfo{volume}{93}},
  \bibinfo{pages}{032302} (\bibinfo{year}{2016}).

\bibitem[{\citenamefont{Sadjadpour et~al.}(2001)\citenamefont{Sadjadpour,
  Sloane, Salehi, and Nebe}}]{sadjadpour_2001}
\bibinfo{author}{\bibfnamefont{H.}~\bibnamefont{Sadjadpour}},
  \bibinfo{author}{\bibfnamefont{N.}~\bibnamefont{Sloane}},
  \bibinfo{author}{\bibfnamefont{M.}~\bibnamefont{Salehi}}, \bibnamefont{and}
  \bibinfo{author}{\bibfnamefont{G.}~\bibnamefont{Nebe}},
  \bibinfo{journal}{Selected Areas in Communications, IEEE Journal on. v19.}
  (\bibinfo{year}{2001}).

\bibitem[{\citenamefont{Vafi and Wysocki}(2005)}]{vafi_2005}
\bibinfo{author}{\bibfnamefont{S.}~\bibnamefont{Vafi}} \bibnamefont{and}
  \bibinfo{author}{\bibfnamefont{T.}~\bibnamefont{Wysocki}}, in
  \emph{\bibinfo{booktitle}{Proceedings of the 6th Australian Communications
  Theory Workshop}} (\bibinfo{year}{2005}).

\bibitem[{\citenamefont{Forney et~al.}(2007)\citenamefont{Forney, Grassl, and
  Guha}}]{forney_2007}
\bibinfo{author}{\bibfnamefont{D.}~\bibnamefont{Forney}},
  \bibinfo{author}{\bibfnamefont{M.}~\bibnamefont{Grassl}}, \bibnamefont{and}
  \bibinfo{author}{\bibfnamefont{S.}~\bibnamefont{Guha}},
  \bibinfo{journal}{IEEE Transactions on information theory}
  \textbf{\bibinfo{volume}{53 - 3}}, \bibinfo{pages}{865}
  (\bibinfo{year}{2007}).

\bibitem[{\citenamefont{Frey and MacKay}(1998)}]{frey_1998}
\bibinfo{author}{\bibfnamefont{B.}~\bibnamefont{Frey}} \bibnamefont{and}
  \bibinfo{author}{\bibfnamefont{D.}~\bibnamefont{MacKay}},
  \bibinfo{journal}{Advances in Neural Infromation Processing Systems}
  (\bibinfo{year}{1998}).

\bibitem[{\citenamefont{Hagiwara and Imai}(2007)}]{hagiwara_2007}
\bibinfo{author}{\bibfnamefont{M.}~\bibnamefont{Hagiwara}} \bibnamefont{and}
  \bibinfo{author}{\bibfnamefont{H.}~\bibnamefont{Imai}}, in
  \emph{\bibinfo{booktitle}{Information Theory. IEEE International Symposium
  on}} (\bibinfo{year}{2007}).

\bibitem[{\citenamefont{Mackay}(1999)}]{mackay_1999}
\bibinfo{author}{\bibfnamefont{D.}~\bibnamefont{Mackay}},
  \bibinfo{journal}{IEEE Transactions on Information Theory}
  \textbf{\bibinfo{volume}{45 No. 2}}, \bibinfo{pages}{399}
  (\bibinfo{year}{1999}).

\end{thebibliography}

\end{document}